\documentclass[structabstract]{aa}

\usepackage{natbib,twoopt}

\usepackage{graphicx,float}

\newcommand{\chandra}{{\it Chandra}}
\newcommand{\swift}{{\it Swift}}
\newcommand{\xmm}{{\it XMM-Newton}}

\usepackage{txfonts}
\begin{document}
   \title{Obscuration effects in super-soft-source X-ray spectra}

   \author{J.-U. Ness\inst{\ref{esa}}\and
  J.P. Osborne\inst{\ref{leicester}}\and
  M. Henze\inst{\ref{esa}}\and
  A. Dobrotka\inst{\ref{slovak}}\and
  J.J. Drake\inst{\ref{sao}}\and
  V. A. R. M. Ribeiro\inst{\ref{capetown}}\and
  S. Starrfield\inst{\ref{asu}}\and
  E. Kuulkers\inst{\ref{esa}}\and
  E. Behar\inst{\ref{haifa}}\and
  M. Hernanz\inst{\ref{barcelona}}\and
  G. Schwarz\inst{\ref{aas}}\and
  K.L. Page\inst{\ref{leicester}}\and
  A.P. Beardmore\inst{\ref{leicester}}\and
  M.F. Bode\inst{\ref{liverpool}}
%
    }

   \institute{Science Operations Division, Science Operations
 	Department of ESA, ESAC, Villanueva de la Ca\~nada (Madrid), Spain;
           \email{juness@sciops.esa.int}\label{esa}
\and
Department of Physics \& Astronomy, University of Leicester, Leicester, LE1 7RH, UK\label{leicester}
        \and
Department of Physics, Institute of Materials Science, Faculty of Materials Science and Technology, Slovak University of Technology in Bratislava, Paulinska 16, 91724 Trnava, Slovak Republic\label{slovak}
        \and
Harvard-Smithsonian Center for Astrophysics, 60 Garden Street, Cambridge, MA 02138, USA\label{sao}
        \and
Astrophysics, Cosmology and Gravity Centre, Department of Astronomy, University of Cape Town, Private Bag X3, Rondebosch 7701, South Africa\label{capetown}
        \and
School of Earth and Space Exploration, Arizona State University, Tempe, AZ 85287-1404, USA\label{asu}
        \and
   Physics Department, Technion, Haifa 32000, Israel\label{haifa}
        \and
Institut de Ci\`encies de l'Espai (CSIC-IEEC), Campus UAB, Facultat de Ci\`encies, C5 parell 2$^{on}$, 08193 Bellaterra (Barcelona), Spain\label{barcelona}
        \and
American Astronomical Society, 2000 Florida Ave., NW, Suite 400,
DC 20009-1231, USA\label{aas}
        \and
   Astrophysics Research Institute, Liverpool John Moores University, IC2 Liverpool Science Park, 146 Brownlow Hill, L3 5RF, UK\label{liverpool}
             }
   \authorrunning{Ness et al.}
   \titlerunning{Obscuration effects in SSS}
   \date{Received \today; accepted }

  \abstract
{Super-soft-source (SSS) X-ray spectra are blackbody-like spectra with
  effective temperatures $\sim 3-7\times 10^5$\,K and luminosities of
  $10^{35-38}$\,erg\,s$^{-1}$.
  Grating spectra of SSS and novae in outburst that show SSS type
  spectra display atmospheric absorption lines. Radiation transport
  atmosphere models can be used to derive physical parameters.
  Blue-shifted absorption lines
  suggest that hydrostatic equilibrium is an insufficient assumption,
  and more sophisticated models are required.
}
   {In this paper, we bypass the complications of spectral models
   and concentrate on the data in a comparative, qualitative
   study. We inspect all available X-ray grating SSS spectra
   to determine systematic, model-independent trends.}
   {We collected all grating spectra of conventional SSS like Cal\,83 and
    Cal\,87 plus observations of novae during their SSS phase. We used
    comparative plots of spectra of different systems to find common
    and different features. The results were interpreted in the context
    of system parameters obtained from the literature.
   }
   {We find two distinct types of SSS spectra that we name SSa and SSe.
    Their main observational characteristics are either clearly visible
    absorption lines or emission lines, respectively, while both types contain
    atmospheric continuum emission. SSa spectra are highly structured
    with no spectral model currently able to reproduce all details.
    The emission lines clearly seen in SSe may also be present in SSa,
    hidden within the forest of complex atmospheric
    absorption and emission features. This suggests that SSe are in fact
    obscured SSa systems.
    Similarities between SSe and SSa with obscured and unobscured
    AGN, respectively, support this interpretation. We find all known or suspected high-inclination systems to
    emit permanently in an SSe state. Some sources are found to
    transition between SSa and SSe states, becoming SSe when fainter.
   }
   {SSS spectra are subject to various occultation processes.
    In persistent SSS spectra such as Cal\,87, the accretion disc
    blocks the central hot source when viewed edge on. In novae during their
    SSS phase, the accretion disc may have been destroyed during
    the initial explosion but could have reformed by the time of
    the SSS phase. In addition, clumpy ejecta may lead to temporary
    obscuration events. The emission lines stem from reprocessed
    emission in the accretion disc, its wind or further out in clumpy
    ejecta,
    while Thomson scattering allows continuum emission to be visible
    also during total obscuration of the central hot source.
    }
   \keywords{novae, cataclysmic variables
 - stars: individual (Cal 87)
 - stars: individual (Cal 83)
 - stars: individual (RXJ 0513-69)
 - stars: individual (T Pyx)
 - stars: individual (Nova LMC 2009a)
 - stars: individual (Nova LMC 2012)
 - stars: individual (V959 Mon)
 - stars: individual (QR And)
 - stars: individual (V4743 Sgr)
 - stars: individual (V2491 Cyg)
 - stars: individual (V723 Cas)
 - stars: individual (RS Oph)
 - stars: individual (KT Eri)
 - stars: individual (U Sco)
 - stars: individual (V1494 Aql)
 - stars: individual (V5116 Sgr)
 - AGN: individual (NGC 1068)
               }

   \maketitle
%

\section{Introduction}

 Supersoft X-ray sources (SSS) are a class of soft X-ray emitters
whose spectra resemble a blackbody in the temperature
range $20-100$\,eV ($3-7\times 10^5$\,K) with luminosities above
$10^{35}$\,erg\,s$^{-1}$ \citep{greiner96}.
The first such sources were found with the {\it Einstein}
Observatory by \cite{cal_discovery}
and were later defined as a class after more were discovered with
ROSAT \citep{truemper92,sssclass,sssclass1}. The shape of the continuum
is caused by atmospheric thermal emission, but
blackbody models were initially used for spectral characterisations.
The first SSS were found in the LMC where X-ray emission is subject to
less photoelectric absorption along the line of sight than in the
case of most Galactic SSS.
In addition to the class of persistent SSS, classical novae (CNe) have
frequently been observed to emit an SSS
spectrum during the later phase of their outburst. It is now commonly
accepted that SSS emission originates in binary
systems containing a white dwarf primary that hosts nuclear burning of
material that is accreted from a secondary star; see
\cite{heuvel} and \cite{kahab}, as well as \cite{darnley_progenitors}
for discussion of types and the role of the secondary star in nova
systems. In persistent SSS, the burning rate is roughly
the same as the accretion rate \citep{heuvel,kahab}, while in novae
during outburst,
the burning rate is higher, eventually leading the SSS emission
to disappear once all hydrogen is consumed.\\

Before a nova explosion occurs, material from the companion is accreted
and accumulates on the white dwarf surface. The pressure and temperature
continuously increase until a thermonuclear runaway occurs, ejecting
a high fraction of the accreted material, together with some core material.
The ejecta initially form an expanding shell surrounding
the white dwarf, which prevents high-energy radiation from escaping.
CNe are optically bright until the density in the outer layers
decreases, exposing hotter layers in the outflowing gases. The
continuing shrinkage of the photosphere ultimately
leads to a shift in the peak of the spectral energy distribution
into the X-ray regime, producing an SSS spectrum when the
photospheric radius is close to a white dwarf radius. SSS emission
declines when nuclear burning ceases once all hydrogen in the
burning zones is either consumed or ejected. For more details we refer
to \cite{st08}.\\

Modelling SSS spectra is challenging, especially for novae.
The SSS spectra observed in low resolution, e.g., with CCDs, can often
be reproduced by a blackbody fit. While no meaningful physical
parameters can be obtained from this approach, it has been shown
that these systems are far from thermal equilibrium; see 
e.g. \cite{balm98}. The physics contained in NLTE atmosphere
models are more realistic, but CCD spectra are not adequate
for constraining LTE nor NLTE parameters because the spectral
resolution of CCDs does not match the complexity of atmosphere models
\citep[e.g.][]{hartheis96,hartheis97}.
\cite{parmarcal83,parmarcal87} have obtained different
results from blackbody fits and NLTE atmosphere models,
but the reproduction of the data is always the same.
\cite{page09} and \cite{osborne11} have obtained similar effective
temperatures from blackbody fits and NLTE atmosphere fits to
\swift/XRT spectra.
The true complexity of radiative transport models has often been
ignored, with only the effective temperature, surface gravity,
and some limited patterns in composition considered as variable
parameters, but the higher
resolution of X-ray grating spectrometers reveals that
this approach is invalid \citep[e.g.][]{sala08,sala10}.\\

With the availability of the grating spectrometers
\xmm/RGS (Reflection Grating Spectrometer), and
\chandra\ LETG/HETG (Low/High Energy Transmission Grating
Spectrometers), simple fitting of atmosphere models with a small
number of parameters is inadequate, and more sophisticated models
are needed.
Currently two approaches are being pursued. Hydrostatic
NLTE models such as TMAP benefit from extensive experience
and a mature understanding of NLTE approximations,
opacities, and atomic physics \citep[e.g.][]{rauch10}.
However, considerable blue shifts of absorption lines
in nova spectra as reported by, e.g., \cite{ness09}
are not reproduced by TMAP, requiring an artificial
correction. Their importance is controversial
because no clear signs of P Cyg profiles have been
found in observed spectra, suggesting a low mass loss
rate.
A self-consistent approach is attempted with the wind
model recently presented by
\cite{vanRossum2012} which is based on spherically symmetric,
expanding, NLTE model atmospheres that have been applied by
\cite{hauschildt92} to novae during their early stages.
The possibility of line absorption caused by expanding
clumps far from the photosphere needs to be tested and
included in the atmosphere calculations if needed.\\

For this work, we bypass the complications of spectral
models and concentrate on the data themselves in a
comparative, qualitative study. Several X-ray grating
spectra of SSS have been obtained, including the two
prototypes \object{Cal\,83} and \object{Cal\,87}
\citep{cal_discovery,cal83_discovery} and a number of
CNe as well as recurrent novae (RNe)
that have been observed during their X-ray bright SSS phase.
We extracted all available archival SSS X-ray grating spectra
to study commonalities between them and group them into two classes.\\

In Sect.~\ref{observations} we describe the sample of targets and
the observations that are discussed in this work. The data are
presented in Sect.~\ref{results} and our interpretations plus
implications of our results are presented in Sect.~\ref{disc}.
We close with our conclusions and a summary in Sect.~\ref{summary}.

\section{Observations}
\label{observations}
 
An overview of all targets discussed in this work is given in
Table~\ref{tab:targets}. We only study SSS spectra, and
exclude all novae that were not observed with an X-ray grating
during their SSS phase. In addition other grating spectra
were extracted for comparison. In Table~\ref{tab:targets}, the sources
are listed by class as defined in the footnotes, date of latest
outburst for novae, system inclination angle if known, and distance.
References for values taken from the
literature are given in the footnotes. Measurements of inclination
angle are not straightforward, yielding rather more secure
measurements for higher inclination angles
via observations of eclipses, illumination effects, or spectroscopic
evidence. Distance measurements
of Galactic sources beyond $\sim 1$\,kpc are highly uncertain as no
direct parallax
measurements of the progenitor systems are available. The upcoming
ESA mission {\it Gaia} will give more precise measurements but
novae that have returned to their quiescent state may be too
faint. Empirical
methods such as maximum magnitude versus rate of decline (MMRD)
relationships are frequently used but are often inaccurate.
For more discussion and a collection of more system parameters,
we refer to \cite{schwarz2011}.\\



\begin{table*}
\begin{flushleft}
\renewcommand{\arraystretch}{1.1}
\caption{\label{tab:targets}Description of targets}
{
\begin{tabular}{lllrr}
\hline
\multicolumn{3}{l}{Target \hfill Type/{\it sub-type}$^a$ \hfill $t_{\rm ref}^b$\ \ \ \ } & $i^c$&$d^d$ \\
\hline
\multicolumn{5}{l}{\bf persistent SSS}\\
Cal\,83         & SSS/SSa    & --    & 20-30$^{[1,2]}$ & 48.1$\pm$2.9$^{[3]}$ \\
Cal\,87         & SSS/\textbf{SSe}    &  --   & 73-77$^{[4,5]}$ & 48.1$\pm$2.9$^{[3]}$\\
RX J0513.9-6951 & SSS/{\it SSa}    &  --   & $\sim$35$^{[6]}$ & 48.1$\pm$2.9$^{[3]}$ \\
QR\,And$^e$     & SSS/{\it SSe}    &  \multicolumn{2}{c}{--\hfill   $35-60^{[7]}$;$>79^{[8,9,10]}$} & 2.0$^{[10]}$ \\
\multicolumn{4}{l}{\bf Classical Novae (CNe)}\\
HV\,Cet$^f$         & CN/{\it SSe}     & 2008-10-07.4 & high$^{[11]}$? & 4.5$\pm$2.0$^{[12]}$ \\
LMC\,2012       & CN/\textbf{SSa}  & 2012-03-26.4 & ? & 48.1$\pm$2.9$^{[3]}$ \\
V959\,Mon       & CN/SSe     & 2012-08-09.8 & 82$\pm$6$^{[16]}$ & 1.5$^{[17]}$ \\
V1494\,Aql      & CN/{\it SSe}     & 1999-12-01.8 & 78.5$^{[18,19,20]}$& 1.6$\pm$0.1$^{[21]}$ \\
V5116\,Sgr      & CN/SSa,e     & 2005-07-04.0 & high$^{[24,26]}$& 11$\pm$3$^{[25]}$\\
V723\,Cas       & CN/{\it SSa}     & 1995-08-24.0 & 62$^{[26,27]}$& 3.9$\pm$0.2$^{[26]}$\\
V382\,Vel       & CN/{\it neb}        & 1999-05-22.4 &  25-67$^{[28]}$ & 1.7$\pm$0.3$^{[29]}$ \\
V4743\,Sgr      & CN/\textbf{SSa}  & 2002-09-19.8 & ? & 3.9$\pm$0.3$^{[15]}$\\
KT\,Eri$^g$     & CN/\textbf{SSa} & 2009-11-14.6 & 58$^{+6}_{-7}{}^{[13]}$& 6.5$^{[14]}$ \\
V2491\,Cyg$^g$  & CN/\textbf{SSa} & 2008-04-10.7 & 80$^{+\ 3}_{-12}{}^{[22]}$?& 10.5$^{[23]}$\\
LMC\,2009a$^g$  & RN/SSe     & 2009-02-05.1 & ? & 48.1$\pm$2.9$^{[3]}$ \\
\multicolumn{5}{l}{\bf Recurrent Novae (RNe)$^b$}\\
RS\,Oph         & RN/\textbf{SSa} & 2006-02-12.8 & 30-40$^{[30,31]}$& 1.6$\pm$0.3$^{[32,33]}$\\
U\,Sco          & RN/\textbf{SSe}  & 2010-01-27.8 & 82$^{[34,35]}$& 12$\pm$2$^{[36]}$\\
T\,Pyx          & RN/{\it neb}        & 2011-04-14.3 & $<20^{[37,38,39]}$& 4.8$\pm$0.5$^{[37,40]}$\\
\hline
\end{tabular}
}

$^a$CN=Classical Nova; RN=Recurrent Nova; SSS=persistent Super Soft Source. Sub-types are given indicating absorption-line dominated X-ray spectra (SSa: Sect.~\ref{sect:sssa}), emission line dominated spectra (SSe: Sect.~\ref{sect:ssse}), and
post-outburst nebular spectra ({\it neb}: Sect.~\ref{sect:neb}). The clearest cases are given in bold face while intermediate or
less clear cases are given in italic. V5116\,Sgr transitions between SSa and SSe, marked by SSa,e (Sect.~\ref{sect:swap}).\\
$^b$Discovery date for CNe and RNe used as reference time (latest outburst for RNe)\\
$^c$Inclination angle in degrees\\
$^d$Distance (kpc); for LMC novae, the distance to the LMC is given \\
$^e$aka RX J0019.8+2156\\
$^f$aka CSS081007:030559+054715\\
$^g$Possible RN\\
$^{[1]}$\cite{cal83_discovery};
$^{[2]}$\cite{schmidtke06};
$^{[3]}$\cite{MSB06};
$^{[4]}$\cite{schandl97};
$^{[5]}$\cite{ebisawa01};
$^{[6]}$\cite{2002AJ....124.2833H};
$^{[7]}$\cite{becker98} ($35<i<60$);
$^{[8]}$\cite{kahab96};
$^{[9]}$\cite{tomov98} ($i>79$);
$^{[10]}$\cite{qrand_discover};
$^{[11]}$\cite{css};
$^{[12]}$\cite{SNO08};
$^{[13]}$\cite{RBD13};
$^{[14]}$\cite{RBS09};
$^{[15]}$\cite{VSS07};
$^{[16]}$\cite{ribeiro13};
$^{[17]}$\cite{2013arXiv1306.1501M};
$^{[18]}$\cite{bos01a};
$^{[19]}$\cite{v1494_eclipse};
$^{[20]}$\cite{hachisu_v1494};
$^{[21]}$\cite{IE03};
$^{[22]}$\cite{ribeiro10};
$^{[23]}$\cite{helton08};
$^{[24]}$\cite{dobrotka_v5116sgr};
$^{[25]}$\cite{sala08};
$^{[26]}$\cite{v723cas_incl};
$^{[27]}$\cite{goranskij07};
$^{[28]}$\cite{shore03};
$^{[29]}$\cite{DPD02};
$^{[30]}$\cite{dobrKen94};
$^{[31]}$\cite{ribeiro09};
$^{[32]}$\cite{B87};
$^{[33]}$\cite{BMS08};
$^{[34]}$\cite{schaefer_eclipse};
$^{[35]}$\cite{thoroughgood01};
$^{[36]}$\cite{schaefer10};
$^{[37]}$\cite{tpyx_pole1};
$^{[38]}$\cite{tpyx_pole2};
$^{[39]}$\cite{SCL13};
$^{[40]}$\cite{2011A&A...533L...8S}.
\renewcommand{\arraystretch}{1}
\end{flushleft}
\end{table*}

\begin{table*}
\begin{flushleft}
\renewcommand{\arraystretch}{1.1}
\caption{\label{tab:obs}Journal of X-ray grating observations}
{
\begin{tabular}{lllrlr|cccc}
\hline
Target & day$^a$ & \multicolumn{2}{l}{ObsID$^b$ \hfill Instrument} & \multicolumn{2}{l}{Start time\hfill Exp.\,time} & CR$^c$ & $f_T^d$ & $f_H^e$ & $f_S^f$\\

&&&&(UT)&(ks)&cts$^{-1}$ & \multicolumn{3}{c}{$10^{-10}$\,erg\,cm$^{-2}$\,s$^{-1}$}\\
\hline
\object{Cal\,83} & & 0077 & LETGS & 1999-11-29.27 & 51.2&\multicolumn{4}{c}{no detection}\\
  & & \multicolumn{2}{l}{0123510101 \hfill RGS} & 2000-04-23.84 & 45.0 & 0.43 & 2.07 & 1.87 & 0.21\\
  & & {\bf 1900} & LETGS & 2001-08-15.67 & 35.2&0.09 & 2.18 & 1.81 & 0.37\\
  & & 3402  & LETGS & 2001-10-03.48 & 61.4 &\multicolumn{4}{c}{no detection}\\
  && \multicolumn{2}{l}{0500860601 \hfill RGS} & 2007-11-24.90 & 22.1 & 0.23 & 2.12 & 1.93 & 0.18\\
\object{Cal\,87}  & & \multicolumn{2}{l}{1896 \hfill ACIS/LETGS} & 2001-08-13.84 & 93.9 &0.03 & 2.20 & 1.94 & 0.26\\
 && \multicolumn{2}{l}{{\bf 0153250101} \hfill RGS} & 2003-04-18.87 & 77.6& 0.12 & 0.61 & 0.32 & 0.28\\
\object{QR\,And}  & & {\bf 0075} & LETGS & 2000-09-28.42 & 51.4&0.01 & 0.99 & 0.65 & 0.35\\
 && \multicolumn{2}{l}{0047940101 \hfill RGS} & 2001-12-31.77 & 56.6 & 0.07 & 0.42 & 0.34 & 0.09\\
\multicolumn{2}{l}{\object{RX J0513.9-6951}} & {\bf 3503} & LETGS & 2003-12-24.29 & 47.7 &0.08 & 1.75 & 1.46 & 0.29\\
&& {\bf 5440} & LETGS & 2005-04-20.80 & 24.6 &0.12 & 2.53 & 2.08 & 0.44\\
&& {\bf 5441} & LETGS & 2005-04-27.95 & 25.0 &0.40 & 5.93 & 5.33 & 0.60\\
&& {\bf 5442} & LETGS & 2005-05-03.24 & 25.5 &0.39 & 5.84 & 5.26 & 0.57\\
&& {\bf 5443} & LETGS & 2005-05-13.82 & 22.5 &0.25 & 4.19 & 3.62 & 0.57\\
&& {\bf 5444} & LETGS & 2005-05-19.09 & 25.0 &0.28 & 4.38 & 3.87 & 0.50\\
&&\multicolumn{3}{l}{{\bf All combined}$^h$} & 170 & 0.23 & 3.40 & 3.17 & 0.23\\
\object{HV\,Cet} & 72.2 & {\bf 9970} & LETGS & 2008-12-18.62 & 34.8 &0.22 & 3.35 & 2.74 & 0.61\\
\object{KT\,Eri} & 71.3   & 12097  & LETGS & 2010-01-23.91 & 14.9 &11.52 & 164 & 163 & 0.76\\
& 79.3 & 12100  & LETGS & 2010-01-31.94 & 27.9 &77.73 & 1031 & 973 & 58.5\\
& 84.6 & 12101  & LETGS & 2010-02-06.27 & 47.8 &37.48 & 509 & 503 & 6.35\\
& 158.8 & {\bf 12203}  & LETGS & 2010-04-21.45 & 32.4 &106.5 & 1564 & 1406 & 158\\
\object{LMC\,2009a}  & 90.4 &\multicolumn{2}{l}{{\bf 0610000301} \hfill RGS} & 2009-05-06.43 & 37.5 & 0.23 & 1.84 & 1.71 & 0.14\\
& 165.0 &\multicolumn{2}{l}{0610000501 \hfill RGS} & 2009-07-20.03 & 56.7 & 1.80 & 16.8 & 14.4 & 2.41\\
& 196.5 & \multicolumn{2}{l}{0604590301 \hfill RGS} & 2009-08-20.59 &14.0 & 2.10 & 16.2 & 13.4 & 2.84\\
& 230.0 &\multicolumn{2}{l}{0604590401 \hfill RGS} & 2009-09-23.02 & 51.0 & 1.23 & 11.0 & 9.42 & 1.56\\
\object{LMC\,2012}  & 31.5 &  {\bf 14426} & LETGS & 2012-04-26.91 & 20.0 &2.12 & 25.7 & 15.4 & 10.3\\
\object{V959\,Mon}  & 116.0 & {\bf 15596} & LETGS & 2012-12-03.82 & 24.9 &0.15 & 2.58 & 1.88 & 0.70\\
\object{V1494\,Aql} & 301.5 & 2308 & LETGS & 2000-09-28.29 & 8.1 &0.72 & 10.5 & 9.22 & 1.26\\
& 304.7 & {\bf 0072} & LETGS & 2000-10-01.42 & 18.1 &0.93 & 12.4 & 11.5 & 0.89\\
\object{V2491\,Cyg} & 39.9 & \multicolumn{2}{l}{{\bf 0552270501} \hfill RGS} & 2008-05-20.59 & 39.2 & 12.6 & 90.1 & 67.2 & 22.9\\
 & 49.6 & \multicolumn{2}{l}{0552270601 \hfill RGS} & 2008-05-30.35 & 29.8 & 2.7 & 17.5 & 12.8 & 4.68\\
\object{V4743\,Sgr} & 180.4 & {\bf 3775} & LETGS & 2003-03-19.40 & 24.7 &38.91 & 540 & 528 & 11.4\\
&196.14 & \multicolumn{2}{l}{0127720501 \hfill RGS} & 2003-04-04.93 & 35.2 & 44.6 & 300 & 289 & 10.7\\
&301.9 & 3776 & LETGS & 2003-07-18.90   & 11.7 &37.16 & 499 & 481 & 18.1\\
&371.0 & 4435 & LETGS & 2003-09-25.99   & 12.0 &19.78 & 276 & 266 & 9.68\\
&526.1 & 5292 & LETGS & 2004-02-28.06   & 10.3 &3.61 & 54.4 & 53.2 & 1.21\\
\object{V5116\,Sgr} & 609.7 & \multicolumn{2}{l}{0405600201 \hfill RGS} & 2007-03-05.72 & 12.8 & 2.1 & 17.5 & 16.4 & 1.09\\
 & 781.8 & 7462 & LETGS & 2007-08-24.80 & 34.8 &0.27 & 4.09 & 3.57 & 0.53\\
\object{V723\,Cas} & 5481.7  & \multicolumn{2}{l}{{\bf 0652070101} \hfill RGS} & 2010-08-26.76 & 50.1 & 0.06 & 0.40 & 0.33 & 0.07\\
 & 6018.1  & \multicolumn{2}{l}{{\bf 0673490101} \hfill RGS} & 2012-02-14.02 & 90.0 & 0.07 & 0.35 & 0.27 & 0.08\\
&&\multicolumn{3}{l}{{\bf All combined}$^h$} & 140 & 0.07 & 0.32 & 0.26 & 0.07\\
\object{RS\,Oph} &26.12 & \multicolumn{2}{l}{0410180201\hfill RGS} & 2006-03-10.96 & 11.7 & 2.5 & 14.8 & 9.89 & 4.92\\
 & 39.7 & {\bf 7296} & LETGS & 2006-03-24.52 & 10.0 &30.16 & 312 & 192 & 120\\
&54.0 & \multicolumn{2}{l}{0410180301 \hfill RGS} & 2006-04-07.88 & 18.6 & 134 & 864 & 457 & 407\\
&66.9 & 7297 & LETGS & 2006-04-20.73 & 6.5 &68.34 & 726 & 408 & 319\\
\object{U\,Sco} & 18.7 & 12102  & LETGS & 2010-02-14.49 & 23.0&0.38 & 5.57 & 4.89 & 0.68\\
 & 22.9 & \multicolumn{2}{l}{0650300201 \hfill RGS} & 2010-02-19.65 & 63.4 & 0.95 & 7.75 & 6.98 & 0.78\\
 & 34.8 &  \multicolumn{2}{l}{{\bf 0561580301} \hfill RGS} & 2010-03-03.61 & 54.3 & 1.1 & 6.74 & 3.35 & 3.40\\
\object{T\,Pyx} & 203.6$^g$ & {\bf 12401} & LETGS & 2011-11-03.86 & 39.8&0.09 & 1.92 & 1.31 & 0.61\\
\object{V382\,Vel} & 267.9$^g$ & 958 & LETGS & 2000-02-14.27 & 24.4 &0.13 & 2.08 & 1.34 & 0.74\\
\object{RS\,Oph} & 111.7$^g$ & 7298 & LETGS & 2006-06-04.50 & 20.0 &0.13 & 2.29 & 1.25 & 1.04\\
\hline
\object{NGC\,1068}&  & \multicolumn{2}{l}{0111200101 \hfill RGS} & 2000-07-29.73 & 42.3 & \multicolumn{4}{l}{For comparison in Fig.~\ref{cmp_rrc}}\\
\hline
\end{tabular}
}

$^a$after $t_{\rm ref}$ (Table~\ref{tab:targets})\\
$^b$Observation Identifiers; those shown in bold indicate observations used for Fig.~\ref{speccat}\\
$^c$Count rate per second\\
$^d$flux over total band 7-38\,\AA; $^e$flux over hard range 7-23\,\AA; $^f$flux over soft range 23-38\,\AA \\
$^g$X-ray observation taken after nova had turned off\\
$^h$For \object{RX J0513.9-6951} and \object{V723\,Cas}, all available observations were combined.\\
\renewcommand{\arraystretch}{1}
\end{flushleft}
\end{table*}

The journal of X-ray observations used in this work is presented in
Table~\ref{tab:obs} with columns: days after discovery for CNe and RNe
relative to the respective reference
times given in Table~\ref{tab:targets}, observation identifier, grating
instrument, start time of observation, and effective exposure time
(after all corrections). The \chandra\ LETGS has been used in
conjunction with the High Resolution Camera (HRC), except for the
observation of \object{Cal\,87}, where the Advanced CCD Imaging Spectrometer
(ACIS) was used as the detector to record the dispersed photons.
For novae, only those observations are included that were
taken during the SSS phase. In the bottom of Table~\ref{tab:obs},
T Pyx is also marked as an observation after the nova had turned
off, along with V382 Vel and RS Oph. Further, we extracted an
\xmm\ spectrum of the obscured Active Galactic Nucleus (AGN)
\object{NGC\,1068} \citep{ngc1068} for comparison.\\

Standard SAS 11.0 and CIAO 3.4 tools were used to obtain calibrated
spectra from the raw \xmm\ and \chandra\ data, respectively. Newer
versions are available but re-extraction is only needed for
quantitative analysis which is not pursued in this work. The count
spectra are constructed by collecting the dispersed photons from
optimised extraction regions. They are then converted to calibrated
flux spectra by dividing the number of counts by the effective area
in each spectral bin which is taken from the calibration of
the respective instruments. For grating spectra, this procedure is
justified because the response matrices are sufficiently
diagonal to yield well-known photon energies for each
spectral channel. While instrumental line broadening
($\sim 0.05$\,\AA) is not corrected, it is not important
for the qualitative analyses performed in this paper.\\

\begin{figure*}[!ht]
\resizebox{\hsize}{!}{\includegraphics{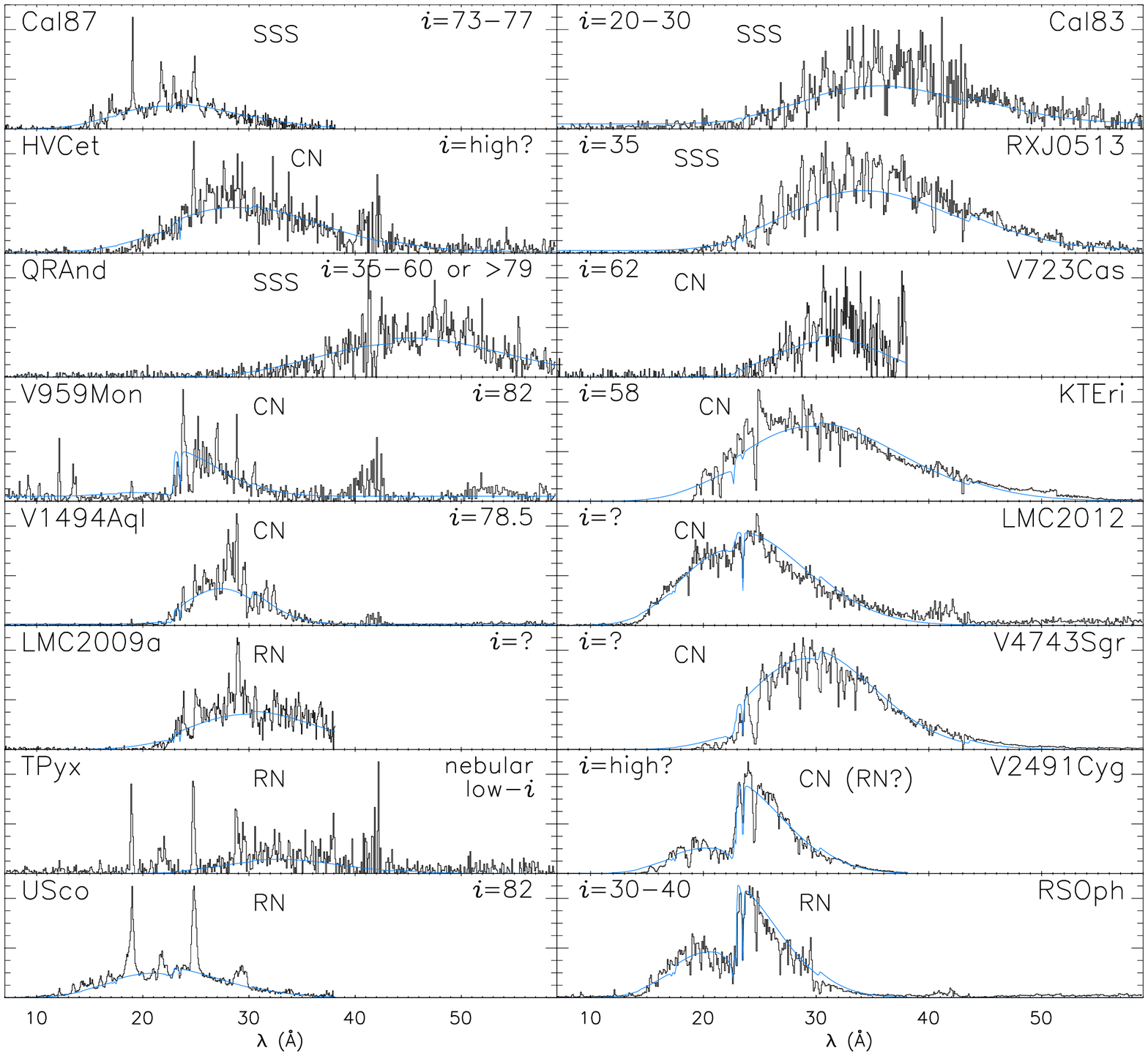}}
\caption{\label{speccat}High-resolution X-ray spectra in
arbitrary flux units of SSS that contain emission
lines without absorption features (SSe, left column) and those that
consist of continuum and absorption lines (SSa, right column).
Inclination angles $i$ from the literature (if known) are given (see
Table~\ref{tab:targets}). The blue thin lines are absorbed blackbody curves,
indicating the presence of photospheric emission in all cases.
The labels SSS, CN, and RN denote persistent supersoft
sources, Classical Novae, and Recurrent Novae, respectively.
An apparent emission feature around 40\,\AA\ in some \chandra\
spectra is an instrumental artefact owing to division by low
effective areas. All six \chandra\ spectra of \object{RX J0513.9-6951}
(second panel in right column) are co-added, and for
\object{V723\,Cas}  (third panel in right column), the two XMM-Newton
spectra have been combined with {\tt rgscombine}.
}
\end{figure*}

Selected calibrated grating spectra are shown in Fig.~\ref{speccat}.
The absorbed blackbody curves are included in Fig.~\ref{speccat_swap}
only to guide the eye to the rough location of the continuum.
Two RGS spectra of \object{V723\,Cas} have been combined using the
SAS task {\tt rgscombine}, and all six \chandra\ spectra of
\object{RX J0513.9-6951} are co-added.\\

\section{Results}
\label{results}

\begin{figure*}[!ht]
\sidecaption
\includegraphics[width=13cm]{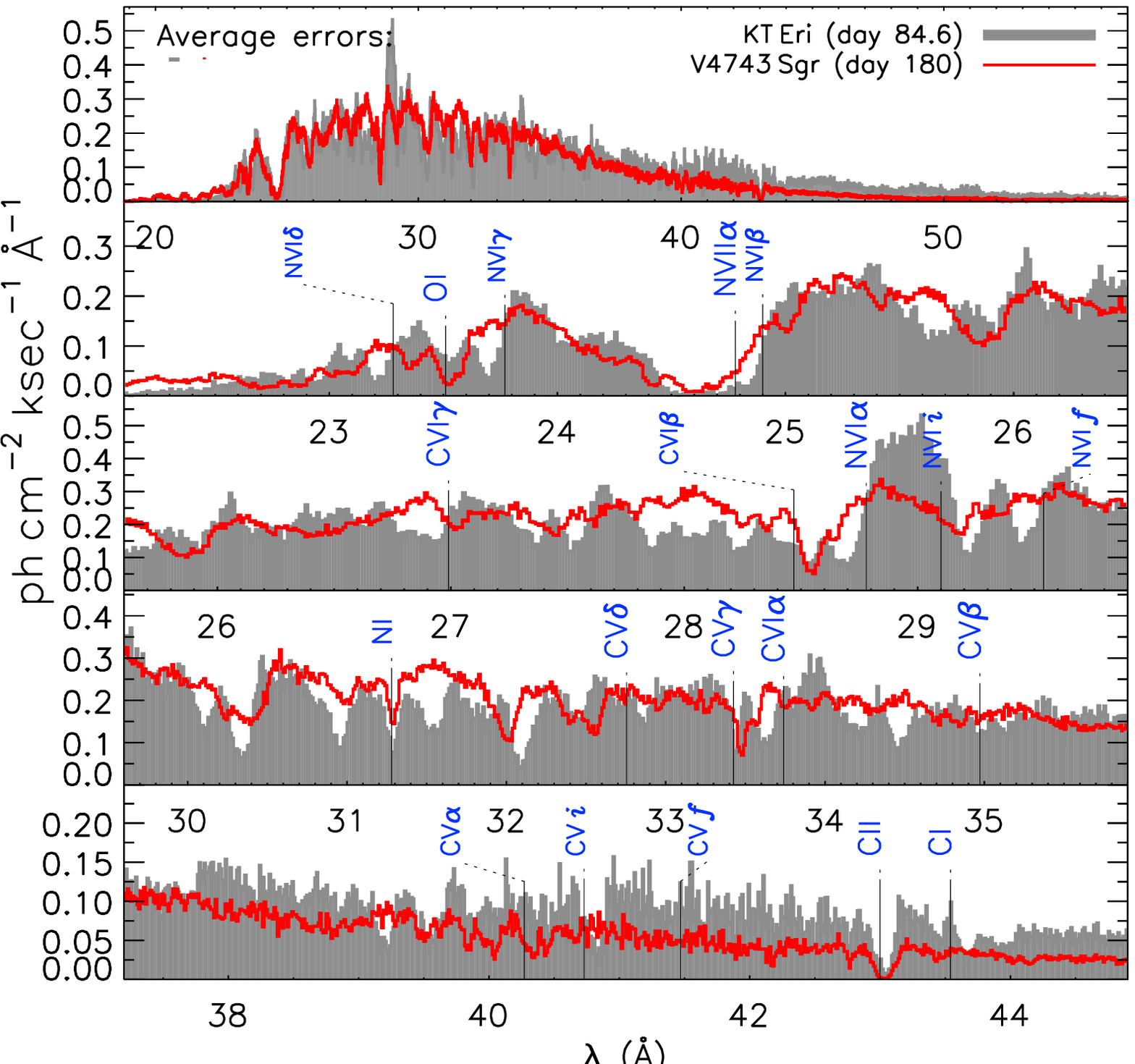}
\caption{\label{cmp_kteriv4743}Comparison plot of grating spectra;
see also Figures \ref{cmp_c87usco}, \ref{cmp_v5116}, \ref{cmp_with_v382},
and \ref{cmp_rrc}.
{\bf Examples of the SSa class}:
Comparison of high-resolution
X-ray spectra of the two CNe \object{KT\,Eri} and \object{V4743\,Sgr}.
No scaling is necessary to fit both spectra in the same graph with the
same units.
\newline
\newline
In this figure as well as in Figures \ref{cmp_c87usco},
\ref{cmp_v5116}, \ref{cmp_with_v382}, and \ref{cmp_rrc},
the top panel shows the entire wavelength range and legends with
average error bars for each observation (left) and line/fill styles
belonging to the corresponding target names (right) where the days in
brackets are days after $t_0$ in Table~\ref{tab:obs}. The panels below
show details, with some overlap spectral ranges. Prominent bound-bound
transitions are marked with vertical lines at their rest wavelengths,
and a label on top of each line gives the corresponding element/ion descriptor.
}
\end{figure*}

Past X-ray observations of SSS have displayed a high degree of
diversity of phenomena, and the objective of this work is to
search for any systematic trends in the data. In light of the
complexity of required models, which unavoidably comes with a high
degree of uncertainty in model parameters, we base our approach
directly on the data. In the following paragraphs, we show and
describe SSS spectra of different systems.

\subsection{Commonalities between SSS spectra}

For comparison of SSS spectra from different systems or
different evolutionary phases, we plot 2-3 spectra in the
same graphs using a combination of lines and shadings of
different colours as marked in the respective legends of
Figures \ref{cmp_kteriv4743}, \ref{cmp_c87usco},
\ref{cmp_v5116}, \ref{cmp_with_v382}, and \ref{cmp_rrc}.
Qualitative comparison of spectra is facilitated by
rescaling if needed, compensating, e.g., for different
intrinsic brightness or distance.
Selected line transitions are labelled at their rest wavelengths
without necessarily claiming the respective lines have been
detected.\\

\cite{ness_v2491} found a remarkable similarity between two SSS
spectra of \object{RS\,Oph} and \object{V2491\,Cyg}, taken on
days 39.7 and 39.9 after their respective outbursts (see their
figure 12). Both sources varied with time
\citep{ness_rsoph,ness_v2491}, and we compared other SSS spectra
of \object{RS\,Oph} and \object{V2491\,Cyg} taken earlier and
later during their respective evolution and found those to be
different.\\

While \cite{ness_v2491} reported fundamentally different spectra
of \object{RS\,Oph} and \object{V4743\,Sgr}, we show in
Fig.~\ref{cmp_kteriv4743} that one of five SSS spectra
of \object{V4743\,Sgr} is in fact remarkably similar to one out
of four SSS spectra of \object{KT\,Eri}; note that no rescaling
is needed!
Both novae were
observed multiple times, and their X-ray spectra varied on
different time scales with \object{V4743\,Sgr} evolving much
slower. Their optical decline time scales show the same
trend, yielding $t_2=9$ and 6.6 days for V4743\,Sgr and
KT\,Eri, respectively \citep{morgan03,hounsell10}. As in
\object{RS\,Oph} and \object{V2491\,Cyg}, both novae evolved differently,
yielding a high degree of similarity only on days 180.4 and
84.6 of \object{V4743\,Sgr} and \object{KT\,Eri}, respectively.
The two pairs of novae \object{RS\,Oph}/\object{V2491\,Cyg}
and \object{V4743\,Sgr}/\object{KT\,Eri} evolved differently
but also passed through an identical spectral phase.\\

\subsection{The SSa subclass}
\label{sect:sssa}

The SSS X-ray spectra of the four systems \object{RS\,Oph}, \object{V2491\,Cyg},
\object{V4743\,Sgr}, and \object{KT\,Eri} are dominated by a blackbody-like
continuum with deep absorption lines (figure~13 in
\citealt{ness_v2491} and Fig.~\ref{cmp_kteriv4743} in this work).
We chose these characteristics to define a new subclass, SSa
(a for absorption lines). After inspection of a Chandra grating
spectrum of the recent nova \object{LMC\,2012}, we also group this nova
into this subclass. In the left column of Fig.~\ref{speccat},
eight examples of SSa spectra are shown.\\

Another object that could be grouped into the SSa subclass
is \object{Cal\,83} but the spectrum is not as well exposed (owing to
the large distance to the LMC).
An NLTE atmosphere model by \cite{lanz04} could be fit to the
\chandra\ spectrum. Also the four SSa spectra of \object{V4743\,Sgr},
\object{V2491\,Cyg}, \object{RS\,Oph}, and \object{KT\,Eri} have been
used to adjust preliminary atmosphere models with some success
(\citealt{rauch10,vanRossum2012,ness_v2491,nelson07})
while for some other observations, spectral modelling has so
far not been successful, e.g. V1494 Aql \citep{rohrbach09}.
Possibly, all SSa spectra can be reproduced by atmosphere
models, in which case \object{Cal\,83} is also an SSa,
while \object{V1494\,Aql} may not be an SSa.\\

With few exceptions, emission lines cannot easily be discerned in
the SSa subclass. The emission lines seen in \object{RS\,Oph} on day 39.7
\citep{ness_rsoph} need to be interpreted in the context of the
cooling shock that occurred between the stellar wind of the
companion and the nova ejecta. \cite{schoenrich07} report
that those lines above the SSS continuum appear enhanced compared
to those at shorter wavelengths, indicating line pumping.
For nova \object{LMC\,2012}, \cite{LMC2012_chan} report detections of emission lines,
and also the SSa spectrum of \object{KT\,Eri} contains additional
emission lines (Fig.~\ref{cmp_kteriv4743}). We consider a strong
possibility that some emission lines are always present in SSS
spectra, albeit not easy to see in SSa spectra.\\

The presence of emission lines complicates spectral modelling
with atmosphere models. More refined dynamic atmosphere models including
mass loss will reveal whether the emission lines are part of
P Cygni profiles (which cannot clearly be decided visually), or whether
they originate from outside of the atmosphere, requiring an independent
model component such as photoionised plasma.\\

{\em In summary, the definition of a member of the SSa class is an X-ray
spectrum that contains a blackbody-like continuum with clearly
visible absorption lines. Additional emission lines may be present
but difficult to see within the complex atmospheric SSa spectrum.}

\subsection{The SSe subclass}
\label{sect:ssse}

\begin{figure*}[!ht]
\sidecaption
\includegraphics[width=13cm]{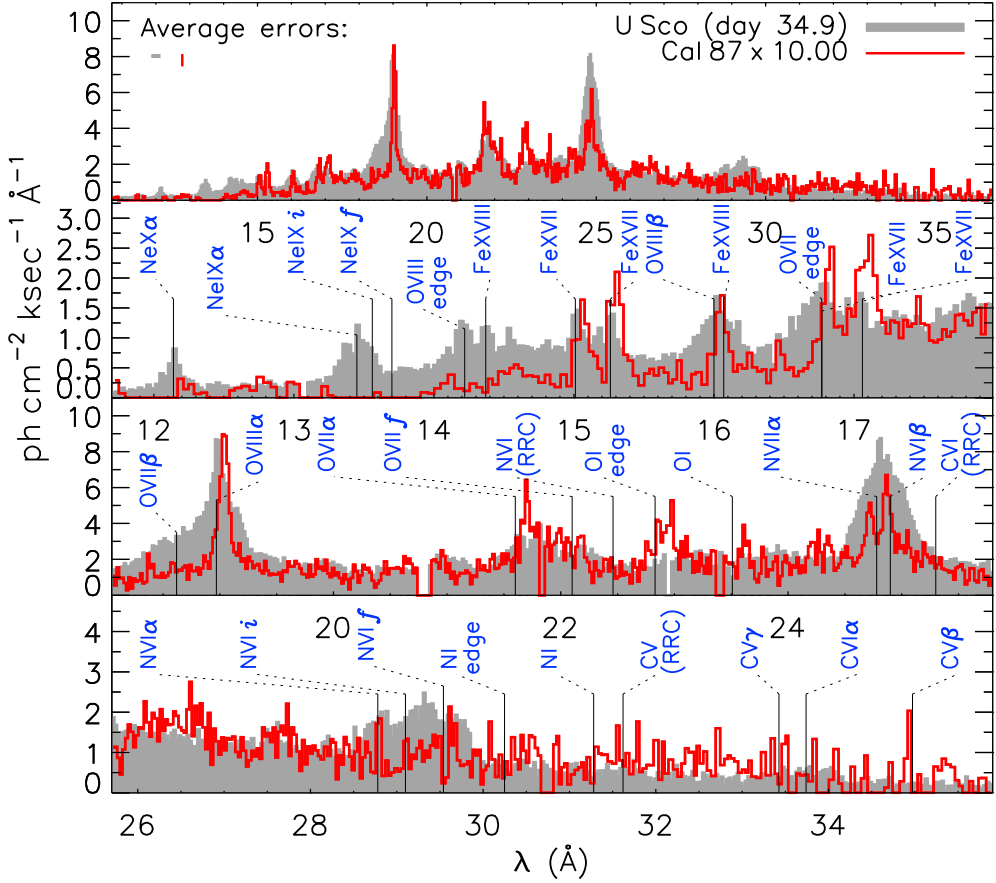}
\caption{\label{cmp_c87usco}
{\bf Examples of the SSe class:}
Comparison of high-resolution X-ray spectra of \object{U\,Sco} and
of the prototype SSS \object{Cal\,87}, scaled by a factor 10 in
brightness.
(The concept is the same as described in Fig.~\ref{cmp_kteriv4743}.)
}
\end{figure*}

During a detailed analysis of \xmm\ observations of \object{U\,Sco},
\cite{ness_usco} noted that this edge-on system resembles the
SSS \object{Cal\,87}. This is demonstrated in Fig.~\ref{cmp_c87usco}
where the \xmm\ grating spectrum of \object{U\,Sco} taken on day 34.9 is
shown in shades of grey and \object{Cal\,87}, scaled by a factor of
10, as a red histogram line.
The strongest lines seen in U\,Sco are also seen
in \object{Cal\,87} but differ in their width. The Ne lines at 12\,\AA\ and
13.5\,\AA\ are only seen in \object{U\,Sco} while several Fe lines are present
in both sources.\\

The strengths of the emission lines seem to be related to the
strength of the underlying continuum at the wavelengths where
they arise. This can be seen in Fig.~\ref{cmp_c87usco} and also
in figure~4 in \cite{ness_usco}.\\

The spectra of these two objects are the clearest examples of
those that differ from the subclass of SSa, and therefore, we
introduce the new subclass SSe, which are SSS spectra that contain
strong emission lines above a weak continuum. The spectra of Cal\,87
and U\,Sco shown in Fig.~\ref{cmp_c87usco} contain no clear signs
of absorption lines.
\cite{ness_usco} concluded from the presence of blackbody-like
continuum emission during times of eclipse in U\,Sco that it results
from Thomson scattering in the electron-rich, ionised nova ejecta.\\

{\em In summary, the definition of a member of the SSe class is an X-ray
spectrum that contains a weak blackbody-like continuum
without absorption lines and exhibits emission lines that
are at least comparable in strength to the continuum. The emission
lines in SSe are strongest where the continuum is strongest.}

\subsection{Intermediate cases}
\label{sect:intermed}

\begin{figure*}[!ht]
\sidecaption
\includegraphics[width=13cm]{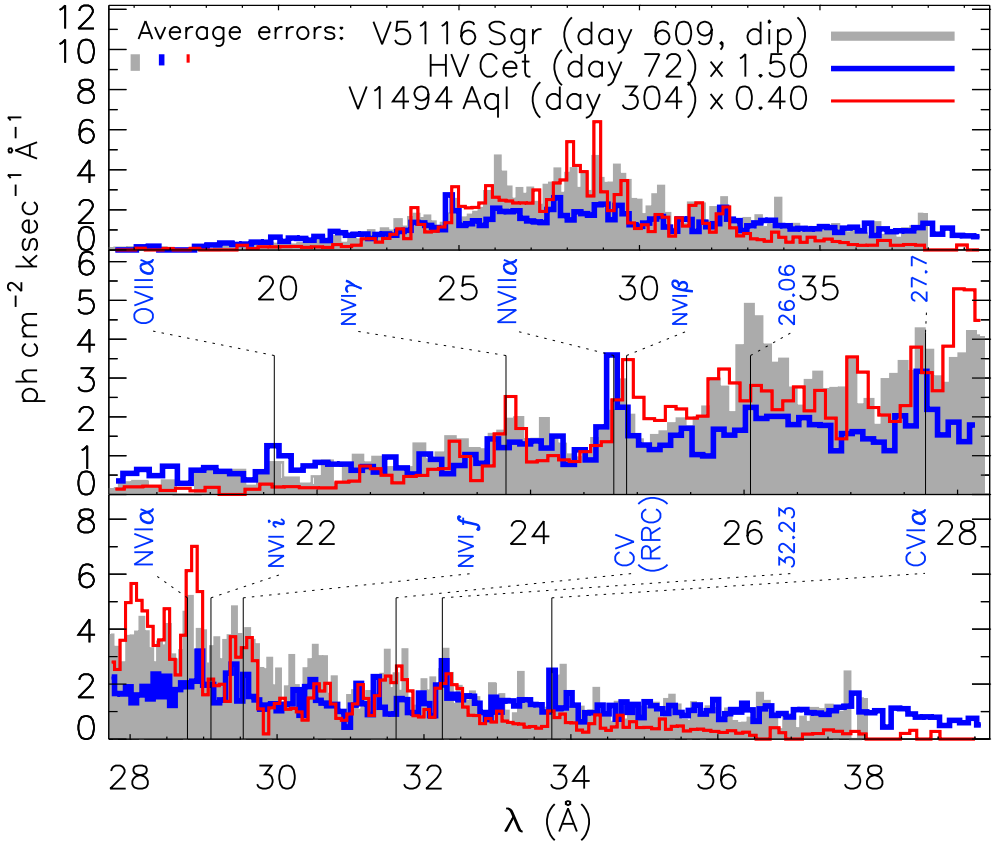}
\caption{\label{cmp_v5116}{\bf Examples of intermediate cases:}
Comparison of high-resolution X-ray spectra of the CNe
\object{V5116\,Sgr}, \object{HV\,Cet}, and \object{V1494\,Aql}.
The wavelengths of
important transitions are marked and labelled. Lines found in other
nova spectra in absorption but without having been identified are
included with their wavelength values.
(The concept is the same as described in Fig.~\ref{cmp_kteriv4743}.)
}
\end{figure*}

The spectral analysis of the grating spectra of the CN \object{V1494\,Aql}
presented by \cite{rohrbach09} turned out to be extremely difficult,
and no satisfactory model was so far found. Some of the nova
X-ray spectra contain a complex mixture of emission- and absorption
lines which is illustrated with three examples in Fig.~\ref{cmp_v5116}.
Emission lines can be seen to occur in all three spectra, including
some unidentified lines that arise at the same wavelengths were absorption
lines were seen in the SSa spectra of \object{V2491\,Cyg},
\object{RS\,Oph}, and \object{V4743\,Sgr} \citep{ness_v2491}.
The three spectra in Fig.~\ref{cmp_v5116} are sufficiently similar
to conclude that these sources belong to the same subclass, and
their characteristics place them more into the SSe subclass, although
these cases are not as clear as \object{U\,Sco} and \object{Cal\,87}
(Fig.~\ref{cmp_c87usco}). Eight examples of SSe are shown in the left
column of Fig.~\ref{speccat}.\\

\subsection{Transitions between SSa and SSe}
\label{sect:swap}

The CN \object{V5116\,Sgr} was highly variable during the \xmm\ observation
taken on day 609.7 \citep{sala08,sala10}.
During an interval of
$\sim 2$\,hours (2/3 of the orbit), the integrated brightness was
more than an order of magnitude fainter than at the beginning of the
observation, and during this time, emission lines were present as
reported by \cite{sala10}.
This could be an example of an SSS
switching between SSa and SSe states, possibly due to a temporary
obscuration event \citep{sala08}.\\

\begin{figure*}[!ht]
\resizebox{\hsize}{!}{\includegraphics{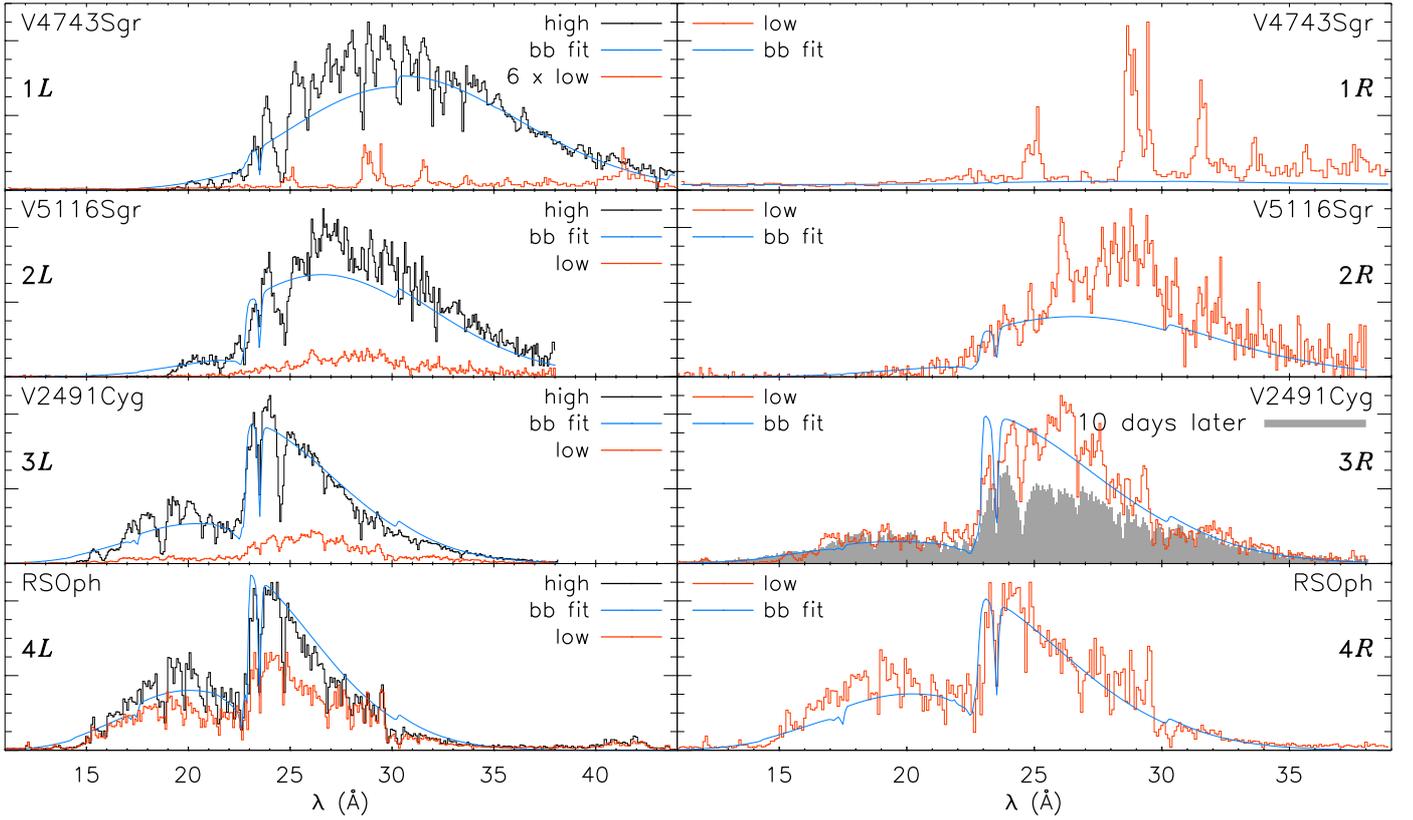}}
\caption{\label{speccat_swap}Transitions between SSa (left) and
SSe (right) states. High-resolution X-ray spectra, in
arbitrary flux units, for four novae with variable light curves.
The left column contains SSa spectra extracted from bright
time intervals while the right column contains SSe spectra
resulting from fainter time intervals.
The light blue lines are absorbed blackbody curves that
guide the eye to the rough location of the continuum. Only
\object{RS\,Oph} (bottom) does not transition from SSa to SSe
during the faint phase.
Panel 3$R$ shows in grey shades the second
XMM-Newton RGS observation of \object{V2491\,Cyg} taken 10 days later,
in comparison with the low-state spectrum. The low-state
spectrum is more of SSe type while the later spectrum is more
of SSa type. We believe that during times of fainter emission,
less continuum is seen, leaving more emission originating further
outside.
}
\end{figure*}

We have studied four novae with variable X-ray light curves
to investigate whether they have transitioned between SSa and
SSe and illustrate the outcome in Fig.~\ref{speccat_swap}.
From the light curves, we extracted spectra
from time intervals of brighter emission (shown in the
left) and fainter time intervals (right). To guide the eye,
blackbody curves are included with blue thin lines.\\

The most extreme case is \object{V4743\,Sgr} (top) in which all
continuum emission was completely obliterated, leaving only
emission lines \citep{v4743}. We speculate that the source of
continuum X-rays either temporarily turned off or was completely
obscured leaving us with emission from only the ejected gas.\\

\object{V2491\,Cyg} was observed twice by XMM-Newton with a
$\sim 3$-hour dip in the first observation (see
figure 10 in \citealt{basi}).
\cite{ness_v2491} subdivided this observation into
three time intervals a, b, c, and in the second row
of Fig.~\ref{speccat_swap}, we show parts c (panel 3$L$)
and b (3$R$). The grey shaded spectrum in panel 3$R$
was taken 10 days later. The
dip spectrum and the later one are about equally bright
but differ decisively between 23-30\,\AA, with the
dip spectrum having some characteristics of an SSe spectrum
while the later spectrum is an SSa spectrum.
The excess in the dip spectrum could have arisen further
outside while the central continuum source was temporarily
obscured.\\

The Chandra grating observation of \object{RS\,Oph} taken on day
39 after outburst was variable (see figure 7 in \citealt{basi}).
This observation was taken during an episode of high-amplitude
variations discovered by \cite{osborne11} in the
\swift\ light curve.
The high- and low-state spectra of \object{RS\,Oph} shown in the
bottom of Fig.~\ref{speccat_swap} are the same as
those shown in figure 9 in \cite{ness_rsoph}. Excess
emission is not as obvious, and this may not be
a transition between SSa and SSe. The similarity of the
brightness and hardness light curves noted by
\cite{ness_rsoph} in their figure 8 suggests
that internal processes (as opposed to external obscuration)
may lead to the variability, perhaps similar to the on- and
off states in \object{RX J0513.9-6951}
\citep{rxj_onoff,burwitz08,southwell96,mcgowan05}
or in \object{Cal\,83} \citep{cal83_firstoff,alcock97,cal83_optcorr}.\\

\subsubsection{Comparison of SSa and SSe spectra}

\begin{figure}[!ht]
\resizebox{\hsize}{!}{\includegraphics{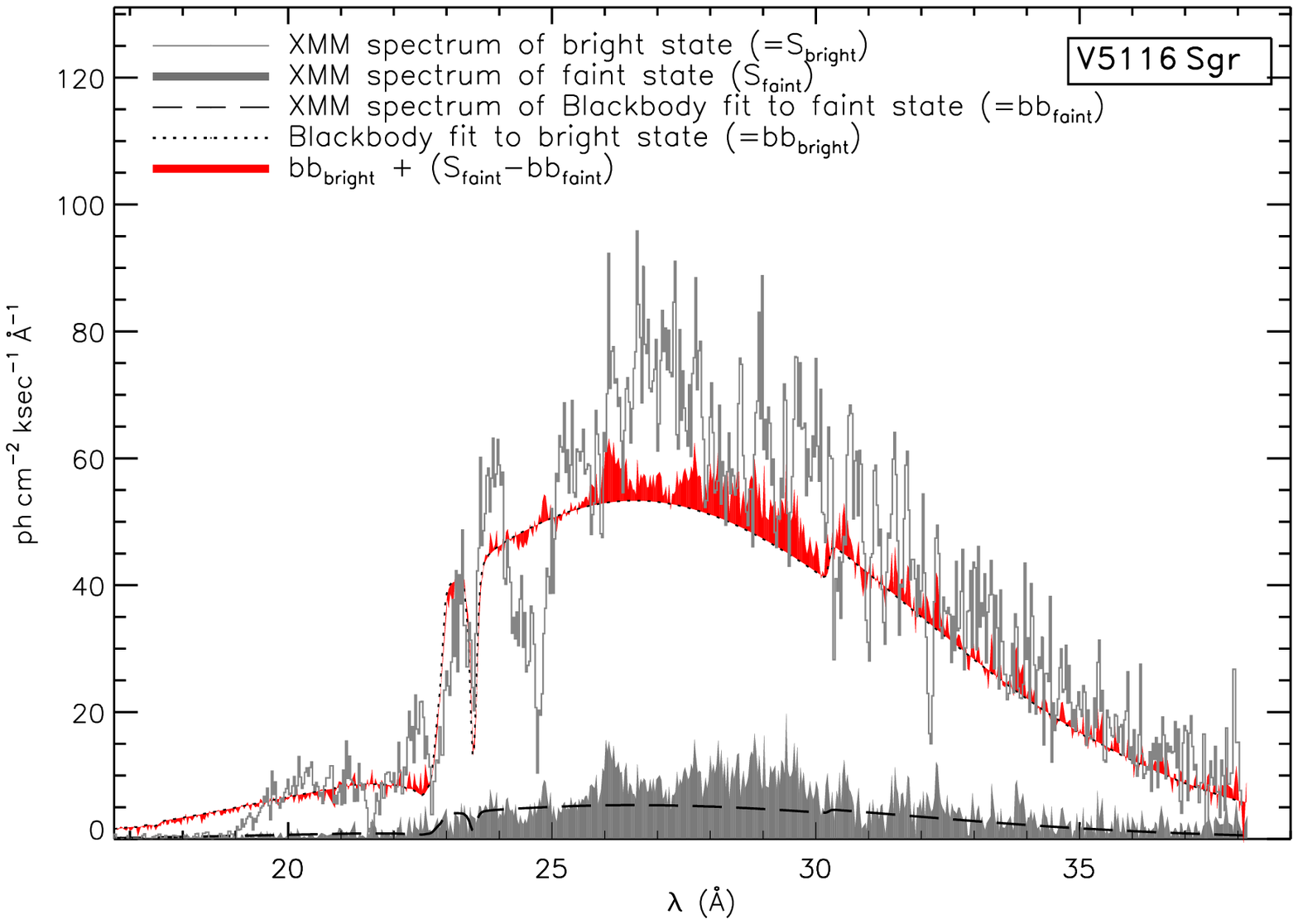}}

\resizebox{\hsize}{!}{\includegraphics{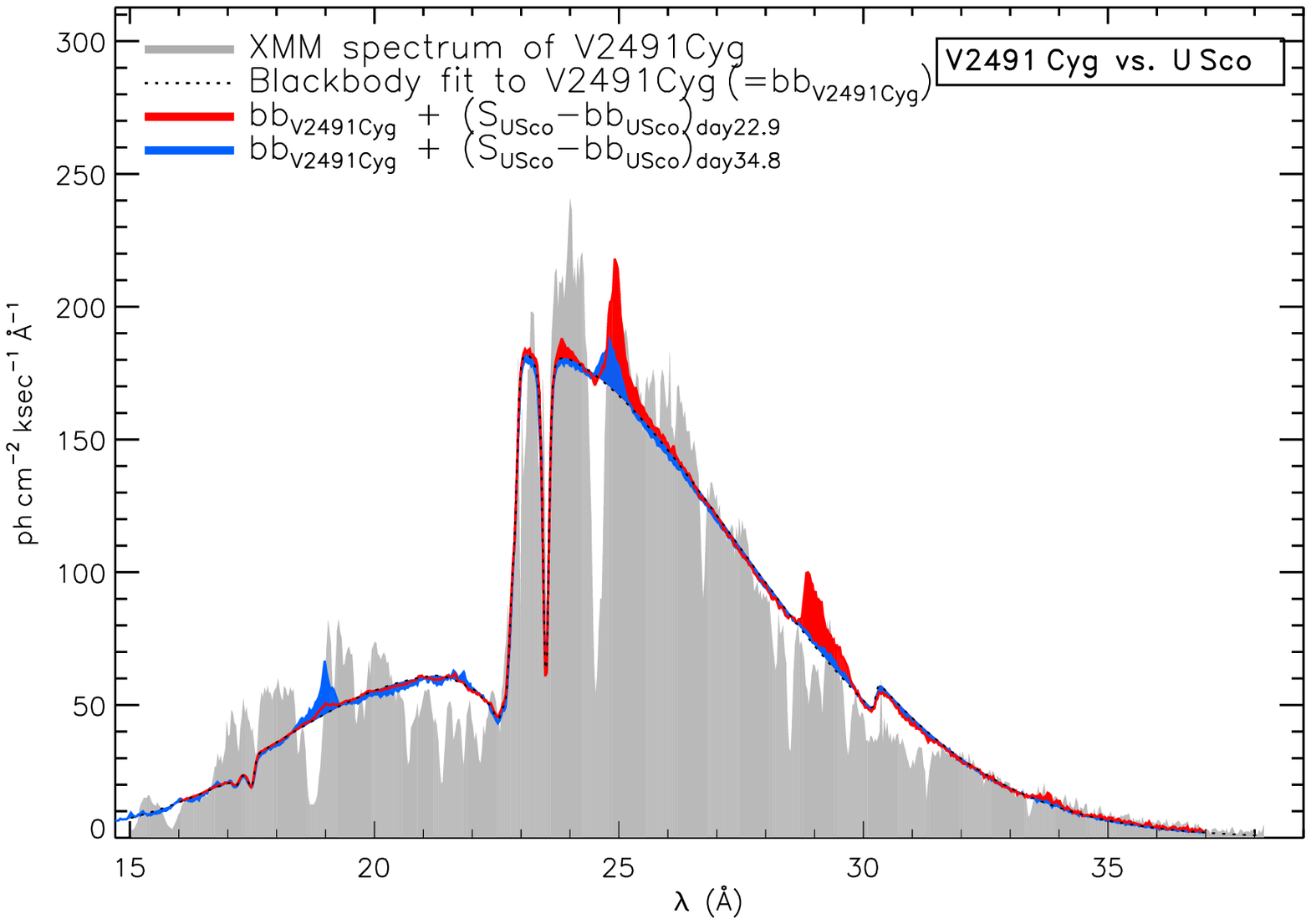}}
\caption{\label{cmp_uscoSSa}Comparison of SSe and SSa
spectra for the high/low state spectra of
\object{V5116\,Sgr} (top) and the two best examples
of SSa, \object{V2491\,Cyg}, and SSe, \object{U\,Sco},
(bottom). The emission line component in SSe spectra,
obtained by removing the weak continuum component, is
added the continuum component of an SSa spectrum.
The continua are approximated by blackbody models.
The SSe emission line components are comparable in
strength to complex atmospheric absorption/emission
features in SSa.
}
\end{figure}

SSS spectra observed in grating resolution contain a multitude
of complex features that no coherent model can currently reproduce.
In order to put the dominant emission lines in SSe spectra
into context with the more complex SSa spectra, we show in
Fig.~\ref{cmp_uscoSSa} direct comparisons between SSa
and SSe spectra. The continuum component of an SSe spectrum is
first estimated via a blackbody fit and then subtracted from
the same observed SSe spectrum. The remaining emission line
component is then added to the continuum component of an SSa
spectrum that was also determined via blackbody estimation.
We performed this operation for the bright/faint spectra of
\object{V5116\,Sgr} (top of Fig.~\ref{cmp_uscoSSa}) and for
the SSa/SSe \object{V2491\,Cyg} and \object{U\,Sco} (bottom
of  Fig.~\ref{cmp_uscoSSa}).\\

For both cases, one can see that the emission lines in SSe
spectra are comparable in strength to various features 
in SSa spectra. Emission lines appearing so dominant in SSe
may thus be present in SSa spectra as well but are hidden
between complex atmospheric features. Without the complex
continuum, the emission line component can be seen much better.\\

\subsection{Pre-SSS and Post-outburst emission lines}
\label{sect:neb}

\begin{figure*}[!ht]
\sidecaption
\includegraphics[width=13cm]{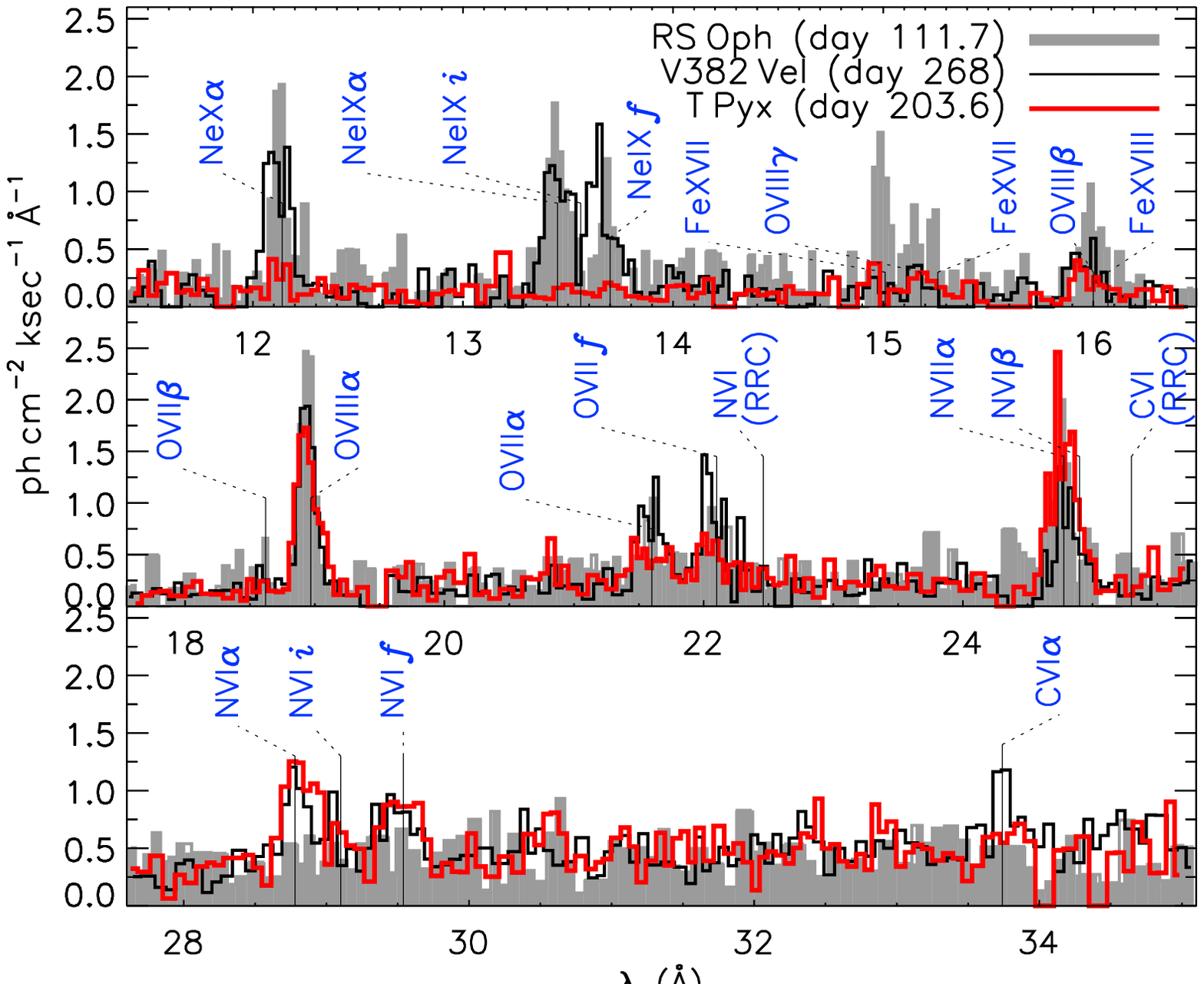}
\caption{\label{cmp_with_v382}Comparison of the post-outburst
high-resolution X-ray spectra of V382\,Vel (think black histogram),
\object{RS\,Oph} (that was an SSa during outburst, thick red histogram),
and the low-inclination system T\,Pyx (grey shadings). The emission
lines in V382\,Vel and \object{RS\,Oph} originate in the
radiatively cooling ejecta after the novae had turned off.
(The concept is the same as described in Fig.~\ref{cmp_kteriv4743}.)
}
\end{figure*}

Emission lines can arise in different contexts. Before the start of
the SSS phase in several novae, emission line spectra had frequently
been observed with CCD spectrometers. While their spectral resolution
is not high enough to resolve emission lines, the nature as emission
line spectra has been established by means of collisional
equilibrium models that reproduce these spectra
\citep[e.g.,][]{krautt96,balm98,mukai01,Orio2001,v458}.
Distinct differences to SSe spectra are:\\
(1) The continuum component in collisional equilibrium originates
from bremsstrahlung, which has a different shape from a blackbody.\\
(2) The emission line fluxes increase in strength during the
SSS phase \citep{schoenrich07}. For the SSe \object{U\,Sco}, only
extremely faint early hard emission was seen \citep{ATel2419}.\\
(3) The emission lines in pre-SSS spectra are in most cases too faint
for detection in grating spectra; exceptions are V959\,Mon
\citep{ATel4569} and symbiotic novae such as RS\,Oph
\citep{rsophshock}.\\
(4) The early hard X-ray component fades with time \citep{v458},
leaving even less potential contributions to SSe spectra.\\

\begin{figure}[!ht]
\resizebox{\hsize}{!}{\includegraphics{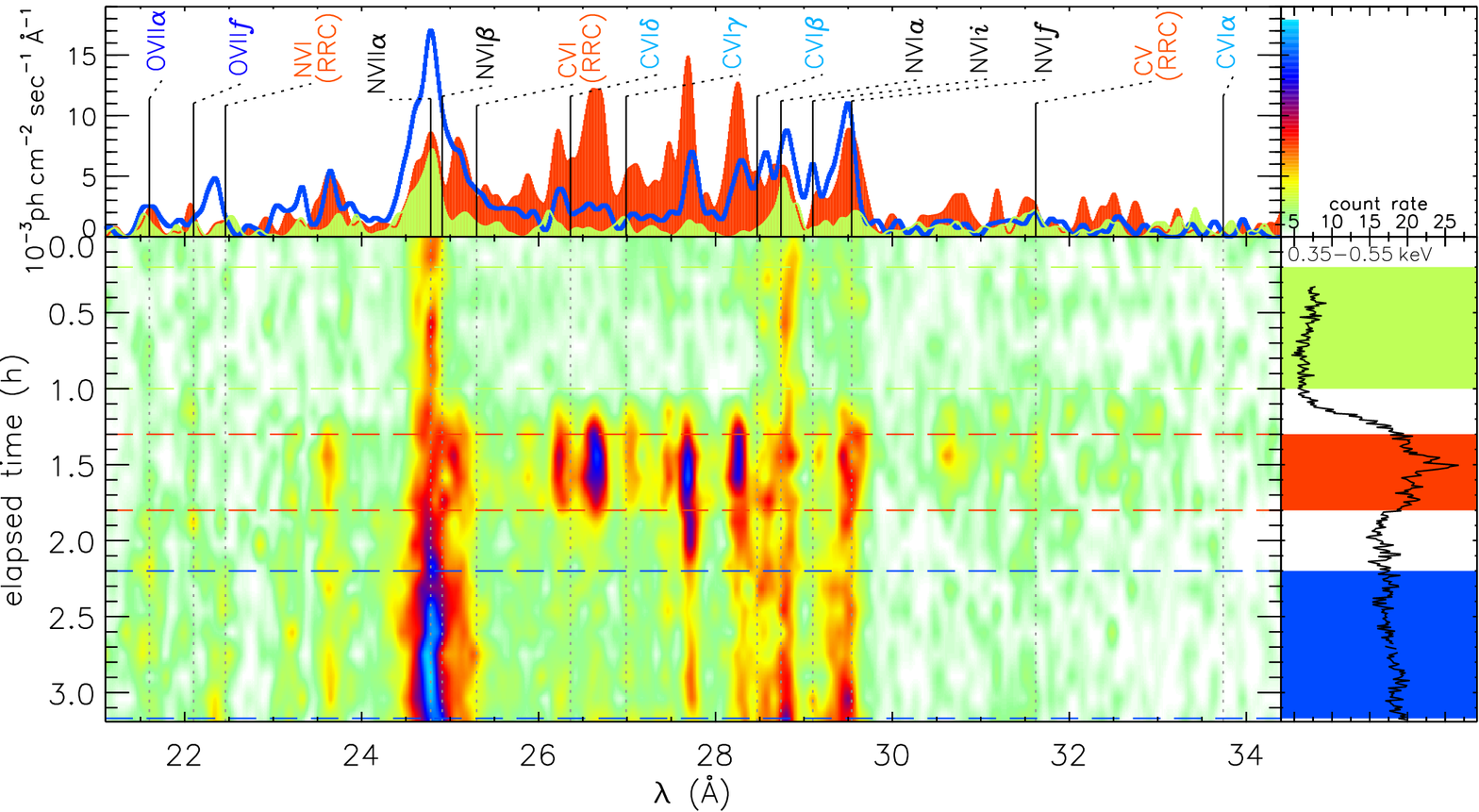}}
\caption{\label{smap_rsoph}Transition into the SSS phase observed
with XMM-Newton in RS\,Oph starting day 26.12 after the 2006 outburst.
The EPIC/pn light curve is shown to the right with three coloured
shades marking time intervals from which the spectra in the top have
been extracted. Line labels that belong to strong transitions in
nitrogen (black), carbon (light blue), and oxygen (dark blue) are included,
as well as Radiative Recombination Continuum (RRC) features in red.
The central image is a brightness spectral/time map
that uses a colour code from light green to light blue representing
increasing flux values as shown with the colour bar along the
spectral flux axis in the top right. See also figure~6 in
\cite{basi}.
}
\end{figure}

The spectral changes with the start of the SSS phase have
serendipitously been observed in an XMM-Newton observation of
RS\,Oph on day 26.12 after the 2006 outburst (see
Table~\ref{tab:obs}) in which \cite{nelson07} discovered the
appearance of a new soft component and showed that this new
component is not yet atmospheric emission from the
white dwarf but a complex emission line spectrum. The time
evolution is illustrated in Fig.~\ref{smap_rsoph}. About one
hour after the start of the observation, an increase in count
rate had occurred that \cite{nelson07} identified as a new soft
component consisting of a complex emission line spectrum. The
transitions belonging to the emission lines between
$\sim 25-29$\,\AA\ are currently unknown, and our spectral time
map shows that they were transient in nature, lasting less than
one hour. At the same time as they disappeared, the N\,{\sc vii}
1s-2p (Ly$\alpha$, 24.78\,\AA) line increased by a substantial
amount, apparently the same amount of flux contained in the
transient emission lines, preserving the integrated count rate
(see right panel). The C\,{\sc vi} Ly$\alpha$ line is not present,
suggesting that none of the transient emission lines originates
from carbon, even if line
shifts are assumed as proposed by \cite{nelson07}. It is possible
that they all arise from nitrogen, although no
transitions of nitrogen are known in this wavelength range.
While a more coherent interpretation of these transient emission
lines is beyond the scope of this paper, the relevant
conclusion for this work is that the pre-SSS emission lines
at soft wavelengths have become significantly stronger with
the start of the SSS phase while harder emission lines have
remained the same; see figure~6 in \cite{basi}. These could
be SSe features that are later hidden between complex atmospheric
structures in SSa spectra.\\

For the four reasons above, we do not consider the emission lines in SSe
as continuation of the early emission lines, and so, in particular
for the persistent SSS \object{Cal\,87}, a different mechanism is
needed. We have restricted the definition of SSe in
Sect.~\ref{sect:ssse} to the presence of emission lines that
coincide in strength with an underlying blackbody-like continuum.\\

Post-outburst X-ray emission line spectra arise in the radiatively
cooling ejecta which has been shown for the post-outburst grating
spectra of V382\,Vel by \cite{ness_vel} and of
\object{RS\,Oph} by \cite{rsophshock}.
These two grating spectra are shown in Fig.~\ref{cmp_with_v382}
in comparison to a \chandra\ spectrum of the RN T\,Pyx, taken
203.6 days after the 2011 outburst \citep{tpyx_chan}. The three
spectra are qualitatively similar, suggesting a similar
origin. Since the \chandra\ spectrum of T\,Pyx was
taken during the decline of the SSS phase, this spectrum
may be a post-outburst spectrum that could arise from radiatively
cooling ejecta.\\

While pre- and post-SSS X-ray spectra are very similar in their
appearance, their origin may still be different. The presence of
high-velocity, dense ejecta during the early evolution of a nova
facilitates shocks with the environment or, in the case of internal
inhomogeneities, shocks within the ejecta. After the SSS phase has
ended, the ejecta have dispersed, leaving only radiative cooling of
optically thin plasma as X-ray production mechanism.
Both are collisional processes, allowing the same spectral models
to be applied, giving only different parameters (such as lower
electron temperature in post-SSS spectra).\\

We do not group post-outburst emission line spectra into the
SSe class because the emission line strengths do not scale
with the strength of the respective underlying SSS continuum.
In V382\,Vel, the SSS spectrum during the active phase did
not exceed 0.6\,keV (20\,\AA) \citep{orio02,burwitz2002a,burwitz2002b}
while in
Fig.~\ref{cmp_with_v382}, emission lines are also seen at
shorter wavelengths. The situation is similar for \object{RS\,Oph},
and in both systems, an early hard component had been present
before the start of the SSS phase \citep{mukai01,rsophshock}.

\subsection{Dependence on Inclination Angle}

The arrangement of the spectra in Fig.~\ref{speccat}
is chosen to display SSe in the left and SSa spectra in the right,
although not all cases can unambiguously be classified (see
Sect.~\ref{sect:intermed}). We include the
inclination angles (if known) from Table~\ref{tab:targets} in the corners
opposite to where the source names are given. As discussed
above, higher inclination angles are more securely determined,
yielding more high values than low values. All inclination angles
above $\sim 70^{\rm o}$ can be found in the left column, indicating
that SSe spectra are a common phenomenon in high-inclination
systems. The only exception may be the SSa \object{V2491\,Cyg} that
was proposed by \cite{ribeiro10} to be a high-inclination
system, however, this has so far not been confirmed by observations
of eclipses.\\

The CN \object{V723\,Cas} is viewed at an intermediate inclination
angle and contains a mixture of SSe and SSa emission, suggesting
that there is a trend with inclination angle. The spectrum
resembles that of \object{QR\,And} which lies at a similar
inclination angle.\\

\section{Discussion}
\label{disc}

Based on the observed sample of SSS spectra, we introduced the two
sub-types SSa and SSe with the primary observational characteristics
being the predominance of absorption lines versus emission lines,
respectively (see definition at end of Sects.~\ref{sect:sssa}
and \ref{sect:ssse}). Some sources transition between SSa and SSe
states, yielding SSe spectra during episodes of fainter emission.
We have shown an influence of inclination angle on the observed
type of SSS spectrum. All systems with known high inclination angle
(which is more securely measurable) display SSe spectra, while for
many SSa spectra, the inclination angle is not known or not
robustly confirmed. Emission lines in SSe spectra appear
comparable in strength to complex sub-structures in SSa spectra,
suggesting that they are always present and can better be
seen when the atmospheric continuum with its complex structure
is not present. SSe spectra
may be the result of obscuration of the continuum component,
exposing emission lines that probably originate from further away
from the white dwarf.\\

\cite{ebisawa01} argued that at the given high inclination angle
in \object{Cal\,87}, the central source was blocked by the accretion disc,
and scattering through the accretion disc corona could explain the
observed spectrum (see also \citealt{ebisawa10}). Similarly,
\cite{css} argue that the apparently sub-Eddington white dwarf
luminosity in \object{HV\,Cet} is due to permanent obscuration of X-rays
by the accretion disc rim. A smooth periodic X-ray modulation
of nova \object{V959\,Mon} \citep{page_mon2012} also makes it likely that
this source is viewed at high inclination and has a white dwarf
surrounded by a scattering region.\\

In the following we discuss the origin of the continuum and the
emission line components (\ref{lineorig}), obscuration effects in
SSe (\ref{obscure}) and resulting implications to
determination of luminosities (\ref{implications}).
Since most SSe spectra originate from high-inclination
systems, the accretion disc likely plays a vital role but the novae
in our sample do not necessarily possess an accretion disc.
We discuss in Sect.~\ref{discreform}
whether the disc has been destroyed during the initial
nova explosion and how it could reform.
Our main conclusion is that SSe are seen when a significant
fraction of the photospheric emission is obscured, but not all X-ray
variability is necessarily due to obscurations
(see Sect.~\ref{othervar}).

\subsection{Origin of the continuum and emission lines in SSe}
\label{lineorig}

The continuum component in all SSS spectra is produced by the
photosphere of the white dwarf. A photospheric temperature of several
$10^5$\,K implies $R\approx 10^{-2}$\,R$_{\odot}$.
 The \object{U\,Sco} system configuration requires the
entire white dwarf be completely eclipsed, but \cite{ness_usco}
found the continuum emission during eclipse centre reduced by
only 50\%.
The only way to see this continuum emission also outside of
the binary orbit is Thomson scattering that preserves the
spectral shape. Thomson scattering can likely also explain
the presence of some if not all continuum emission in SSe.\\

The origin of the emission lines on top of a blackbody-like
continuum in SSe spectra is less clear. As discussed in
Sect.~\ref{sect:neb}, we do not consider the emission lines
in SSe as a continuation of early emission lines sometimes
seen in novae.\\

The strongest emission lines in SSe arise at wavelengths where
the continuum is strongest. This is not typical of collisional
equilibrium spectra and is rather suggestive of photoexcitation
which should not be confused with photo{\it ionisation} that
has been disfavoured by \cite{orio_usco} for U\,Sco. Ions in
excited states re-emit previously absorbed photons in all directions,
independently of the location of the ionising source. Continuum
photons emitted from the white dwarf into polar directions can
effectively be scattered into other directions via this process.
Because it is most efficient for resonance lines with
high oscillator strengths, it is called resonant line scattering
and has been proposed for \object{Cal\,87} by \cite{greiner_cal87} and
for \object{U\,Sco} by \cite{ness_usco}. Depending on the
geometry of the scattering medium, photons can be scattered into
the line of sight or out of the line of sight. Since these processes
balance out in a spherically symmetric scattering medium, it needs
to be asymmetric to allow for observable effects of resonant line
scattering.\\

The contribution of resonant scattering to the emission lines
dominates that of recombination when the column density through
the absorber/re-emitter is low, but decreases as the column
increases \citep{ngc1068}. The detection of Radiative Recombination
Continuum (RRC\footnote{RRCs are emission features reflecting the
released energy during recombination.}) features is a sign of
recombination, but note that RRCs can be detected only
for low temperatures. If there is additional heating (on top of
photo-heating), which brings the gas to higher temperatures
(e.g., exceeding $\sim 30$\,eV, and dependent on spectrometer used
and S/N), the RRCs would be broadened in the spectrum beyond
recognition. In that case, the plasma would be more collisional.\\

If the lines are the result of a mixture of processes, then it 
is difficult to explore them for quantitative diagnostics.
In the case of pure resonance scattering, the line intensity 
is a measure of how the scattering medium covers the source.
Based on the atomic cross section for the resonant
Ly$\alpha$\footnote{1s-2p transitions in H-like ions such as
O\,{\sc viii} or C\,{\sc vi} all have the same oscillator strength.}
transitions,
for significant line absorption and scattering, an ionic column
density of at least $10^{15}$\,cm$^{-2}$ is needed (figure 7 of
\citealt{ngc1068}), with some dependency on line broadening.
Hence, if all observed Ly$\alpha$ emission is attributed to
resonant scattering, the H column is greater than
$\sim 10^{18}$\,cm$^{-2}$, assuming cosmic abundances.
The significance of a discrete RRC emission feature rises as the
column density of the scattering medium increases. In particular,
the flux ratio of the RRC to Ly$\alpha$
rises from $\sim 0.1$ to 1.0 as the column density increases from
$10^{16}$\,cm$^{-2}$ to $10^{18}$\,cm$^{-2}$. This is demonstrated
using an atomic-state kinetics model in figures 1 and 2 of
\cite{ngc1068Behar}. If we can estimate the luminosity and spectrum 
of the central source when it is obscured, we could get a rough idea 
of the covering fraction of the emitting gas.\\

For recombination or collisional line excitation the emission measure
formalism can be applied to derive the product of density square and
volume, $n_e^2V$, and independent measurement of either quantity gives
the other.\\

\subsection{Obscuration of the continuum in SSe}
\label{obscure}

Independent of the formation mechanisms of the emission lines,
the decisive difference between SSa and SSe is the brightness
of the continuum relative to the strengths of emission lines
which are about the same in both subclasses. The most likely
explanation for the weaker continuum emission in SSe is
obscuration of the central source while some continuum emission
can escape obscuration via Thomson scattering and reprocessing.\\

\subsubsection{Comparison to obscured and unobscured AGN}
\label{sect:agn}

Obscuration of the continuum with remaining predominant emission lines
as in our SSe spectra has also been observed in some Active Galactic
Nuclei (AGN). Indeed, this is the central concept of the unified model
of AGN, in which the spectrum observed is strongly dependant on whether
the accretion structures of the central supermassive black hole are
observed at low or high inclination, leading to a view which is
unobscured (type I) or obscured (type II) by a dusty torus at large
radii respectively \citep{agnunify}.
For example, \cite{ngc1068,brinkman2002} describe such an emission
line spectrum from \object{NGC\,1068} and conclude that it is produced by
photoionisation and photoexcitation, followed by recombination
and de-excitation, respectively, either directly into the ground
state or into lower excited states from where they cascade into the
ground state. An AGN spectrum equivalent to an SSa is that of the
unobscured AGN NGC 3783.
\cite{ngc3783} report the detection of 135 absorption lines, 42
of which are unblended. In addition, they report
line emission ''filling in'' the absorption, while the combination
of absorption and emission lines is not interpreted as a P Cygni
profiles. Their figure~1 around 19\,\AA\ looks like the SSS spectrum
of \object{KT\,Eri} around 24\,\AA\ with one single emission line
clearly detectable above the continuum (Fig.~\ref{cmp_kteriv4743}).
In the case of NGC 3783, there are clear emission lines at
longer wavelengths where there is no continuum which can
be explained by photoexcitation by the high-energy continuum.
On the other hand, emission lines belonging to transitions
with energies higher than the Wien tail of an SSS continuum
can not be photo-excited.
Detailed analysis of a mixture of continuum emission and
emission lines seen in \object{NGC\,1365} (figure~2 in \citealt{ngc1365})
reveals a mixture of collisional and photo-ionisation.\\

Important features in the X-ray spectra of obscured AGN
arise from RRC features which \cite{ngc1365} attribute to a
photoionised plasma. Their width
depends on the mean electron velocity, or kinetic electron
temperature. RRCs can only be observed in a plasma of
electron temperatures up to some 100\,eV, not sufficient
to reach the ionisation stages of observed lines such as
O\,{\sc viii} via collisional ionisation. The presence
of narrow RRC features is a strong indicator for a cool
plasma that reaches high ionisation stages via photoionisation.

The inverse process, absorption during ionisation, leads to
absorption edges in the ionising continuum spectrum at the
same energies as RRCs. In perfect ionisation equilibrium,
neither RRC nor ionisation edges are observable. In
obscured AGN, the ionisation source cannot be seen, and
the recombination emission from the photoionised plasma
is not balanced by the continuum emission with absorption
edges from the source.

\begin{figure}[!ht]
\resizebox{\hsize}{!}{\includegraphics{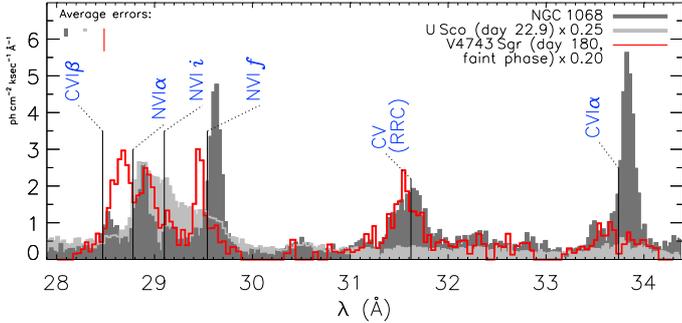}}
\caption{\label{cmp_rrc}Comparison of an \xmm/RGS spectrum of
AGN \object{NGC\,1068} (dark shades) with \object{U\,Sco} (light shades)
and the last 5.5\,ks of the March 2003 \chandra\ observation of
\object{V4743\,Sgr} (thick red line; see Table~\ref{tab:obs}).
\object{NGC\,1068} contains strong radiative recombination features
(RRC; \citealt{ngc1068}). The C\,{\sc v} RRC at 31.5\,\AA\
can also be seen in \object{V4743\,Sgr} but not in \object{U\,Sco}.
(The concept is the same as described in Fig.~\ref{cmp_kteriv4743}.)
}
\end{figure}

The only nova spectrum with RRC features was taken during
the intermittent faint state in \object{V4743\,Sgr} reported
by \cite{v4743}. In Fig.~\ref{cmp_rrc}, we compare an \xmm\
spectrum of the AGN \object{NGC\,1068} with \object{U\,Sco}
and the faint spectrum of \object{V4743\,Sgr} with scaling
factors applied as given in the legend. Apart from a scaling
factor, the C\,{\sc v} RRC feature at 31.5\,\AA\ is identical
in \object{NGC\,1068} and \object{V4743\,Sgr}. The same width indicates
similar electron temperatures. Also the strength in relation
to the N\,{\sc vi} lines is remarkably similar, however, the
N\,{\sc vi} lines are red shifted in \object{NGC\,1068} and
blue shifted in \object{V4743\,Sgr} while the C\,{\sc v} RRC
feature resides at the rest wavelength.
The spectrum of \object{U\,Sco} contains no RRC features
\citep{orio_usco}, suggesting a different formation origin.

\subsubsection{Approximation of luminosity in SSa and SSe}
\label{implications}

To illustrate the effect of continuum obscuration further, we
compare the bolometric luminosities derived for SSa and SSe
spectra. To avoid the ambiguity from the large parameter space
of atmosphere models, we focus here
on blackbody fits to illustrate the effects.
Obtaining well constrained, credible parameters
from atmosphere models is not the purpose of this project.\\

\begin{figure}[!ht]
\resizebox{\hsize}{!}{\includegraphics{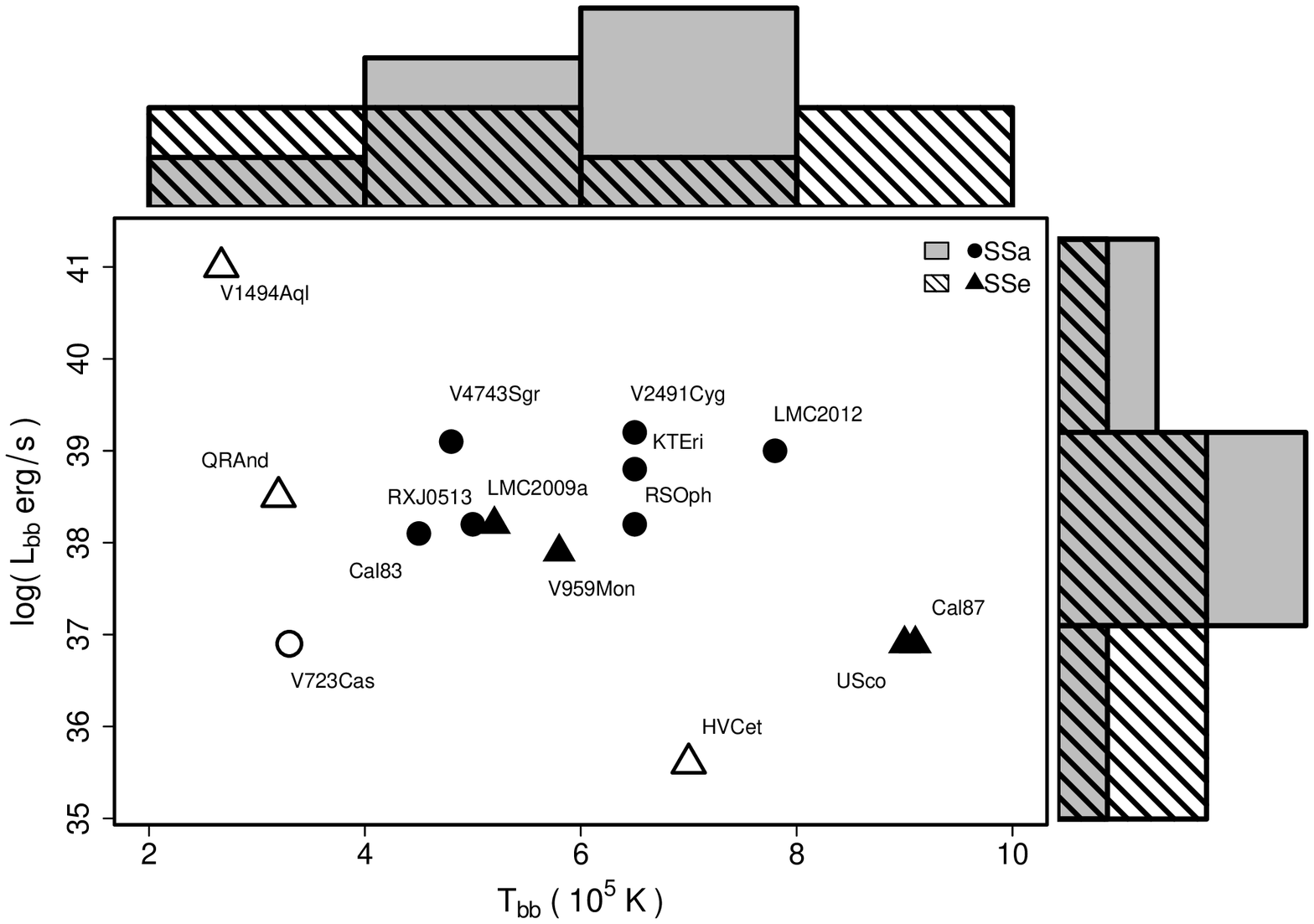}}

\resizebox{\hsize}{!}{\includegraphics{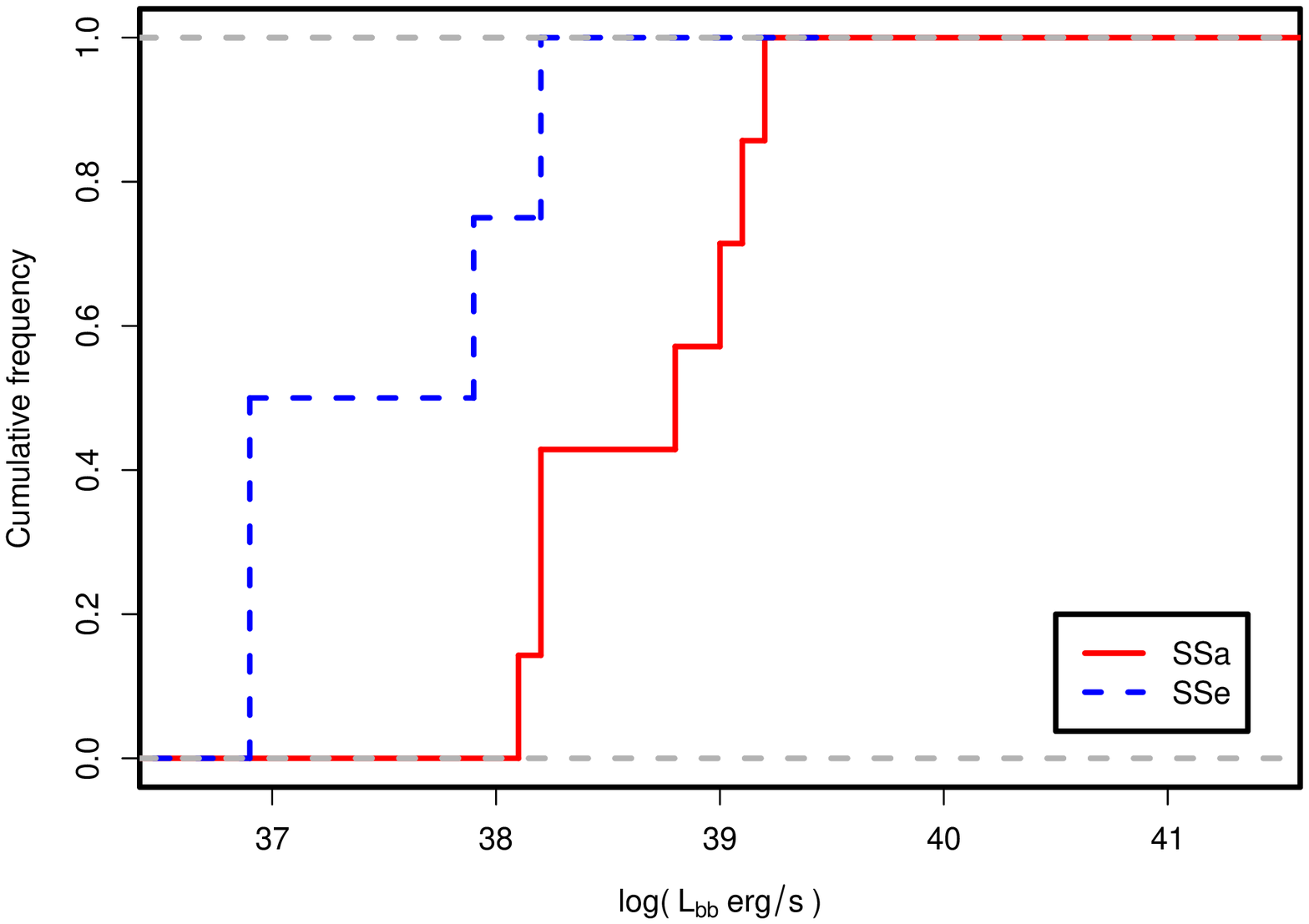}}
\caption{\label{nh_lum}{\bf Top}: Blackbody-estimated luminosities,
$L_{\rm bb}$, versus temperature, $T_{\rm bb}$, determined from
Planck functions fitted to the continuum component of SSa and SSe
spectra, marked by bullet and triangular symbols, respectively.
Open symbols indicate particularly high degree of uncertainty
owed to line blends or high but poorly constrained amounts of
photoelectric absorption.
The hatched and shaded histograms in the top and right,
decoded in the legend, give the number of SSe and SSa spectra
in rough temperature and luminosity intervals.
{\bf Bottom}: Cumulative distributions of derived
luminosities for SSa and SSe, including only the systems
with filled symbols in the top panel. SSa clearly
yield higher luminosities.
BB-derived luminosities and temperatures are indicative,
as they are not the result of a full-physics modelling of
the data, and for this reason error bars are not plotted.
}
\end{figure}

We use the parameters of the light blue blackbody curves
included in Fig.~\ref{speccat}, and in the top panel of
Fig.~\ref{nh_lum}, we
show the values of $L_{\rm bb}$ versus $T_{\rm bb}$.
It is well known that blackbody fits can yield
unrealistic luminosities \citep{krautt96} and
consequently also temperatures. We use the values
as indicators without giving error bars, as they are
not the result of a full-physics modelling of the data.
Atmosphere models have more flexibility to stay below
the Eddington luminosity (see, e.g., \citealt{balm98,page09}),
but there is no unique way of accomplishing that, and the
multitude of possibilities may obscure the point we are aiming
at. Major sources of uncertainty are the distance (values used are
given in Table~\ref{tab:obs}) and contamination by
blending absorption and/or emission lines that complicate the
determination of an appropriate model which we caution again
was only done manually.
In addition, sources with high values
of $N_{\rm H}$ suffer from a high degree of systematic uncertainty
as a large fraction of the radiation from the source is actually
not visible to us and thus not available to constrain models
that correspond to different values in bolometric luminosity and
effective temperature.\\

The results for the new subclasses SSa and SSe are plotted
with bullet and triangular symbols, respectively.
In the top and right of the plot, distribution histograms
are shown indicating that the distribution in temperature and
luminosity yields higher values for SSa spectra.
Clearly, the sample is too small to draw reliable conclusions,
and more sources are desirable to populate the diagram.\\

In the bottom panel of Fig.~\ref{nh_lum}, cumulative
distributions are shown for the luminosities derived
from SSe (blue) and SSa (red) that have more secure
values. The distributions clearly show that systematically
lower luminosities are derived for SSe spectra.
A Wilcoxon rank sum test rejects the null hypothesis
that both luminosity distributions agree on the
95\% confidence level, thus supporting the conclusion
that the central source is obscured in SSe compared to
SSa.

\subsection{Effects from an accretion disc}
\label{discreform}

We found a systematic trend of high-inclination systems
displaying SSe spectra with no confirmed case of an SSa
spectrum emitted by a high-inclination system. The most
obvious reason for obscuration of the central continuum
in high-inclination systems is the accretion disc if present.
The phenomenon of predominance of emission lines over weak
continuum in high-inclination systems has also been observed
dwarf novae (DNe) in outburst
\citep{hamilton07}, revealing similar conclusions drawn from
the X-ray grating spectrum of \object{U\,Sco} \citep{ness_usco}.
\cite{kuulkers_review} summarise that in DNe, eclipses affect
the continuum more strongly than the lines which is
consistent with our observation that SSe generally have
relatively weaker continuum than SSa. Eclipsing DNe
such as WZ\,Sge and OY\,Car are dominated more by emission
lines than systems with lower inclination angle such as
SS\,Cyg \citep{mauche04}.
\cite{MaucheRaymond2000} found good spectral fits to EUVE
spectra of OY Carinae using a model similar to obscured AGN
(=Seyfert 2 galaxies) wherein the EUV radiation from the central
regions and the disc is scattered into the line of sight by the
system's photo-ionised disc wind. Similar processes were
also observed in Low Mass X-ray Binaries (LMXB) which contain a
neutron star that accretes via an accretion disc from a
low-mass companion star. \cite{kallman03} presented
X-ray grating spectra of the high-inclination LMXB
2S 0921-63
and also pointed out the analogous nature to obscured
AGN as we do in Sect.~\ref{sect:agn}.\\

To determine the limiting angle above which the central source
is obscured by the accretion disc, we consider the ratio of the
disc radius $R_{\rm d}$ and the outer disc edge $H_{\rm d}$ which can be
estimated as a function of $M_1$, $M_2$ (primary and secondary
masses, respectively) and orbital period. A standard thin $\alpha$-disc
approximation \citep{FrankKingRaine92}
yields $H_{\rm d}/R_{\rm d} \sim 0.01$, thus $H_{\rm d}$ is small compared to
$R_{\rm d}$ (and also to the white dwarf), allowing occultation
only at extremely high inclination
angles. However, there are additional processes that lead to
a thicker outer edge of the disc such as the impact of the accretion
stream from the secondary star (e.g., \citealt{frank87,armitage98}, invoked by \citealt{page_mon2012})
or the presence of the hot central SSS source irradiating the
disc at a level of several
$10^5$\,K \citep{suleimanov99,suleimanov03}. Accounting for
the resulting "spray" at the outer disc edge, $H_{\rm d}/R_{\rm d}$ can
reach values from 0.25 to 0.4 (see figure~3 in \citealt{schandl97}),
thus yielding large disc heights compared to the radius of
the central source. The spray can then eclipse the white dwarf also at
lower inclination angle. From simple trigonometry, the arctangent
of the $H/R$ ratio gives the angle between the disc plane and the white
dwarf-spray line. It is the angle below which the white dwarf is
obscured by the spray (assuming the white dwarf being 
a point source). The resulting inclination ($90^{\rm o} -$ this angle) is 
approximately $68^{\rm o}$ for $H/R = 0.4$ and $76^{\rm o}$ for
$H/R = 0.25$. The central source can then be obscured by the spray
above inclination, in the range from $68^{\rm o} - 76^{\rm o}$\footnote{We
note that, if the accretion disc is tilted or warped, the lower limit
on the limiting angle can be even lower.}.
From trigonometric considerations, we find that for inclination
angles below $68^{\rm o}$, the white dwarf
is unobscured, yielding an SSa spectrum, while above $76^{\rm o}$, a
pure SSe spectrum is expected, which is confirmed by
Fig.~\ref{speccat}. Between these two values, the white dwarf
is partially eclipsed, yielding hybrid spectra between SSa and
SSe. Since the spray is unlikely to be uniform in height above the
disc, transitions between SSa and SSe may vary with the orbit,
however, this can not be confirmed from the observations so far.\\

While an accretion disc is present in persistent SSS, CVs in
general, and in DNe, it might be absent in novae during their
SSS phase. The
initial explosion destroys the accretion disc. For
example for \object{U\,Sco}, \cite{drake10} performed
hydrodynamic simulations of the initial blast concluding that
the accretion disc was completely destroyed.
For the novae
in our sample, the disc may have either re-formed before our
spectra were obtained, or it was never destroyed in the first place
as proposed by, e.g., \cite{walter11}.\\

First signs of re-establishment of the accretion disc in \object{U\,Sco}
were renewed optical flickering some 7 days after outburst
\citep{IAUC9114,munari10}. While a time scale of 7 days
would be consistent with predictions of the mass loss rate from
the companion by \cite{hachisu00}, any renewed disc build-up
needs to overcome the dynamic pressure of the radiatively driven
outflow from the evolving central object at this early time
\citep{drake10}. Meanwhile, the X-ray studies during the
later SSS phase have shown that an accretion disc may have
reformed by the start of the SSS phase \citep{ness_usco}.
At a later time, the wind from the central object may have
weakened. Furthermore, a disc may reform more rapidly after
a nova explosion because of the heating effect on the secondary
star that would cause enhanced Roche lobe overflow
\citep[e.g.][]{heuvel}.\\

\subsection{Other sources of X-ray variability}
\label{othervar}

The transitions between SSa and SSe were not seen during
fainter episodes of \object{RS\,Oph} (see Fig.~\ref{speccat_swap}),
and this system has in fact shown a correlation between
brightness and hardness, lagging by 1000\,seconds
\citep{ness_rsoph,schoenrich07}. Associated changes in spectral
shape suggest a physical origin such as photospheric
expansion, leading to a decrease in brightness and
shift of the SED to longer wavelengths at which the
radiation is absorbed by the interstellar medium
\citep{swnovae}.\\

Also,
the decline in brightness in \object{V4743\,Sgr} has been
accompanied with a softening of the X-ray spectrum
\citep{v4743}, and it is also the only nova showing
RRC features during a faint state (Fig.~\ref{cmp_rrc}).
Recently, \cite{cal83_optcorr}
have studied X-ray on- and off states in \object{Cal\,83}
which are inversely correlated with optical high- and
low states. They found a softening of the X-ray
spectrum when the source increases in optical
brightness. Perhaps the underlying processes to
this type of variability are related to the
high-amplitude variations frequently seen in novae
during the early SSS phase. While the discussion of
variability deserves close inspection, this is
beyond the scope of this paper.

\section{Summary and Conclusions}
\label{summary}

Before \xmm\ and \chandra\ revolutionised X-ray astronomy,
the observed SSS spectra seen as plain blackbodies with no
features in them that could be used to sensibly constrain
any more sophisticated models. The X-ray gratings that can
resolve lines are required to explore SSS spectra to deeper
limits. These limits have not yet been reached by
current-generation models, and we focus only on the
data.\\

Over the last 13 years, high-resolution grating spectra of
SSS and various types of novae during their SSS phase have
been obtained. Here, we studied all available grating spectra
and found remarkable similarities between some of them, even
when taken from different systems at different times since
outburst.
While some SSS spectra exhibit emission from a photosphere
with absorption lines, others are dominated by
emission lines on top of a weak blackbody-like continuum.
We define two SSS classes SSa
and SSe for the former and latter types, respectively:
\begin{itemize}
 \item The SSa class shows an X-ray spectrum of a 
  blackbody-like continuum with clearly visible absorption
  lines. Emission lines may be present but are difficult to see.
 \item The SSe class shows an X-ray spectrum that is clearly
 dominated by emission lines from well-known species.
 They also contain a weak blackbody-like continuum
 of comparable intensities to the emission lines but without clear
 absorption lines.
\end{itemize}
We found two pairs of almost identical SSa spectra,
RS\,Oph/V2491\,Cyg and V4743\,Sgr/KT\,Eri,
and one pair of identical SSe spectra, U\,Sco/Cal\,87.
In addition, various intermediate spectra
have been observed that contain characteristics of
both.\\

All the evidence we gathered together points to an interpretation
of obscuration of the central source in SSe spectra
compared to SSa spectra:

\begin{itemize}
 \item Systems with high inclination angles $>70^{\rm o}$
   are seen to have SSe spectra.
 \item In some novae, we find both SSa and SSe states,
   where the SSe state was always seen during episodes
   of reduced brightness, indicating obscuration of the
   continuum component.
 \item If the white dwarf is completely blocked, e.g. by the
   companion, blackbody-like continuum emission may still
   be observable because of Thomson scattering and reprocessing.
 \item We observed stronger emission lines in SSe at wavelengths
   where the underlying continuum is stronger, indicating
   photoexcitations as part of resonant line scattering.
 \item SSa spectra may contain emission lines of the same
   strengths as in SSe which are hidden between complex
   atmospheric features.
 \item Blackbody fits to the continuum component yield
   systematically lower luminosities in SSe than in SSa.
 \item Similar phenomena are known for other types of CVs,
    dwarf novae. Resemblance of an SSe spectrum with that
    of an obscured AGN supports an interpretation that
    central continuum emission is obscured in SSe.
\end{itemize}

Permanent obscuration in high-inclination angle systems implies
the presence of an accretion disc. In nova systems, the disc
may either have survived the initial explosion or it has
reformed by the time of the start of the SSS phase.\\

We caution, however, that not all observed variability
need be attributed to obscuration. The high amplitude oscillations
in \object{RS Oph} described by \cite{osborne11} have not shown the
characteristic fading to an SSe spectrum, but associated hardness
changes. Likewise, recent studies of
on- and off states in \object{Cal\,83} point to physical effects
as hardness changes also occur during transitions between
on- and off states. These effects are not the subject of
this work but deserve closer inspection.

\begin{acknowledgements}
We thank the referee for a most inspiring and encouraging
report. The authors appreciate useful discussions with M. Giustini about
AGNs. We thank S.N. Shore who has helped with useful comments.
This research has made use of data obtained with the gratings
on board \xmm\ and \chandra. \xmm\ is an ESA
science mission with instruments and contributions directly funded by
ESA Member States and NASA.
Software provided by the \chandra\ X-ray Center (CXC) in the
application package CIAO was used to obtain science data.
K.L. Page, J.P. Osborne, and A.P. Beardmore acknowledge financial support from the UK Space Agency.
A. Dobrotka was supported by the Slovak Academy of Sciences Grant No. 1/0511/13 and by the ESA international fellowship.
M. Henze acknowledges support from an ESA fellowship.
V.A.R.M. Ribeiro acknowledges the South African SKA Project for funding the postdoctoral fellowship at the University of Cape Town.
M. Hernanz acknowledges the Spanish MICINN grant AYA2011-24704 and FEDER funds.
SS acknowledges partial support from NASA and NSF grants to ASU.
 \end{acknowledgements}


\bibliographystyle{aa}
\bibliography{cn,astron,jn,rsoph,usco,vel,valerio}

\begin{thebibliography}{119}
\expandafter\ifx\csname natexlab\endcsname\relax\def\natexlab#1{#1}\fi

\bibitem[{{Alcock} {et~al.}(1997){Alcock}, {Allsman}, {Alves}, {Axelrod},
  {Bennett}, {Charles}, {Cook}, {Freeman}, {Griest}, {Guern}, {Lehner},
  {Livio}, {Marshall}, {Peterson}, {Pratt}, {Quinn}, {Rodgers}, {Southwell},
  {Stubbs}, \& {Sutherland}}]{alcock97}
{Alcock}, C., {Allsman}, R.~A., {Alves}, D., {et~al.} 1997, MNRAS, 286, 483

\bibitem[{{Antonucci} \& {Miller}(1985)}]{agnunify}
{Antonucci}, R.~R.~J. \& {Miller}, J.~S. 1985, ApJ, 297, 621

\bibitem[{{Armitage} \& {Livio}(1998)}]{armitage98}
{Armitage}, P.~J. \& {Livio}, M. 1998, ApJ, 493, 898

\bibitem[{{Balman} {et~al.}(1998){Balman}, {Krautter}, \& {\"Ogelman}}]{balm98}
{Balman}, S., {Krautter}, J., \& {\"Ogelman}, H. 1998, ApJ, 499, 395

\bibitem[{{Barry} {et~al.}(2008){Barry}, {Mukai}, {Sokoloski}, {Danchi},
  {Hachisu}, {Evans}, {Gehrz}, \& {Mikolajewska}}]{BMS08}
{Barry}, R.~K., {Mukai}, K., {Sokoloski}, J.~L., {et~al.} 2008, in Astronomical
  Society of the Pacific Conference Series, Vol. 401, RS Ophiuchi (2006) and
  the Recurrent Nova Phenomenon, ed. A.~{Evans}, M.~F. {Bode}, T.~J. {O'Brien},
  \& M.~J. {Darnley}, 52

\bibitem[{{Barsukova} \& {Goranskii}(2003)}]{v1494_eclipse}
{Barsukova}, E.~A. \& {Goranskii}, V.~P. 2003, Astronomy Letters, 29, 195

\bibitem[{{Beardmore} {et~al.}(2012){Beardmore}, {Osborne}, {Page}, {Hakala},
  {Schwarz}, {Rauch}, {Balman}, {Evans}, {Goad}, {Ness}, {Starrfield}, \&
  {Wagner}}]{css}
{Beardmore}, A.~P., {Osborne}, J.~P., {Page}, K.~L., {et~al.} 2012, \aap, 545,
  A116

\bibitem[{{Becker} {et~al.}(1998){Becker}, {Remillard}, {Rappaport}, \&
  {McClintock}}]{becker98}
{Becker}, C.~M., {Remillard}, R.~A., {Rappaport}, S.~A., \& {McClintock}, J.~E.
  1998, ApJ, 506, 880

\bibitem[{{Behar} {et~al.}(2002){Behar}, {Kinkhabwala}, {Sako}, {Paerels},
  {Kahn}, {Brinkman}, {Kaastra}, \& {van der Meer}}]{ngc1068Behar}
{Behar}, E., {Kinkhabwala}, A., {Sako}, M., {et~al.} 2002, in Astronomical
  Society of the Pacific Conference Series, Vol. 255, Mass Outflow in Active
  Galactic Nuclei: New Perspectives, ed. D.~M. {Crenshaw}, S.~B. {Kraemer}, \&
  I.~M. {George}, 43

\bibitem[{{Beuermann} {et~al.}(1995){Beuermann}, {Reinsch}, {Barwig},
  {Burwitz}, {de Martino}, {Mantel}, {Pakull}, {Robinson}, {Schwope}, {Thomas},
  {Truemper}, {van Teeseling}, \& {Zhang}}]{qrand_discover}
{Beuermann}, K., {Reinsch}, K., {Barwig}, H., {et~al.} 1995, A\&A, 294, L1

\bibitem[{{Bode}(1987)}]{B87}
{Bode}, M.~F. 1987, in RS Ophiuchi (1985) and the Recurrent Nova Phenomenon,
  ed. M.~F. {Bode}, 241

\bibitem[{{Bos} {et~al.}(2001){Bos}, {Retter}, {McCormick}, \&
  {Velthuis}}]{bos01a}
{Bos}, M., {Retter}, A., {McCormick}, J., \& {Velthuis}, F. 2001, IAUcirc,
  7610, 2

\bibitem[{{Brinkman} {et~al.}(2002){Brinkman}, {Kaastra}, {van der Meer},
  {Kinkhabwala}, {Behar}, {Kahn}, {Paerels}, \& {Sako}}]{brinkman2002}
{Brinkman}, A.~C., {Kaastra}, J.~S., {van der Meer}, R.~L.~J., {et~al.} 2002,
  A\&A, 396, 761

\bibitem[{{Burwitz} {et~al.}(2008){Burwitz}, {Reinsch}, {Greiner},
  {Meyer-Hofmeister}, {Meyer}, {Walter}, \& {Mennickent}}]{burwitz08}
{Burwitz}, V., {Reinsch}, K., {Greiner}, J., {et~al.} 2008, A\&A, 481, 193

\bibitem[{{Burwitz} {et~al.}(2002{\natexlab{a}}){Burwitz}, {Starrfield},
  {Krautter}, \& {Ness}}]{burwitz2002b}
{Burwitz}, V., {Starrfield}, S., {Krautter}, J., \& {Ness}, J.-U.
  2002{\natexlab{a}}, in American Institute of Physics Conference Series, Vol.
  637, Classical Nova Explosions, ed. M.~{Hernanz} \& J.~{Jos{\'e}}, 377--380

\bibitem[{{Burwitz} {et~al.}(2002{\natexlab{b}}){Burwitz}, {Starrfield},
  {Krautter}, \& {Ness}}]{burwitz2002a}
{Burwitz}, V., {Starrfield}, S., {Krautter}, J., \& {Ness}, J.-U.
  2002{\natexlab{b}}, in High Resolution X-ray Spectroscopy with XMM-Newton and
  Chandra, ed. G.~{Branduardi-Raymont}

\bibitem[{{Chesneau} {et~al.}(2011){Chesneau}, {Meilland}, {Banerjee}, {Le
  Bouquin}, {McAlister}, {Millour}, {Ridgway}, {Spang}, {Ten Brummelaar},
  {Wittkowski}, {Ashok}, {Benisty}, {Berger}, {Boyajian}, {Farrington},
  {Goldfinger}, {Merand}, {Nardetto}, {Petrov}, {Rivinius}, {Schaefer},
  {Touhami}, \& {Zins}}]{tpyx_pole1}
{Chesneau}, O., {Meilland}, A., {Banerjee}, D.~P.~K., {et~al.} 2011, A\&A, 534,
  L11

\bibitem[{{Crampton} {et~al.}(1987){Crampton}, {Cowley}, {Hutchings},
  {Schmidtke}, {Thompson}, \& {Liebert}}]{cal83_discovery}
{Crampton}, D., {Cowley}, A.~P., {Hutchings}, J.~B., {et~al.} 1987, ApJ, 321,
  745

\bibitem[{{Darnley} {et~al.}(2012){Darnley}, {Ribeiro}, {Bode}, {Hounsell}, \&
  {Williams}}]{darnley_progenitors}
{Darnley}, M.~J., {Ribeiro}, V.~A.~R.~M., {Bode}, M.~F., {Hounsell}, R.~A., \&
  {Williams}, R.~P. 2012, ApJ, 746, 61

\bibitem[{{Della Valle} {et~al.}(2002){Della Valle}, {Pasquini}, {Daou}, \&
  {Williams}}]{DPD02}
{Della Valle}, M., {Pasquini}, L., {Daou}, D., \& {Williams}, R.~E. 2002, A\&A,
  390, 155

\bibitem[{{Dobrotka} {et~al.}(2008){Dobrotka}, {Retter}, \&
  {Liu}}]{dobrotka_v5116sgr}
{Dobrotka}, A., {Retter}, A., \& {Liu}, A. 2008, A\&A, 478, 815

\bibitem[{{Dobrzycka} \& {Kenyon}(1994)}]{dobrKen94}
{Dobrzycka}, D. \& {Kenyon}, S.~J. 1994, AJ, 108, 2259

\bibitem[{{Drake} \& {Orlando}(2010)}]{drake10}
{Drake}, J.~J. \& {Orlando}, S. 2010, ApJL, 720, 195

\bibitem[{{Ebisawa} {et~al.}(2001){Ebisawa}, {Mukai}, {Kotani}, {Asai},
  {Dotani}, {Nagase}, {Hartmann}, {Heise}, {Kahabka}, \& {van
  Teeseling}}]{ebisawa01}
{Ebisawa}, K., {Mukai}, K., {Kotani}, T., {et~al.} 2001, ApJ, 550, 1007

\bibitem[{{Ebisawa} {et~al.}(2010){Ebisawa}, {Rauch}, \& {Takei}}]{ebisawa10}
{Ebisawa}, K., {Rauch}, T., \& {Takei}, D. 2010, AN, 331, 152

\bibitem[{{Frank} {et~al.}(1992){Frank}, {King}, \& {Raine}}]{FrankKingRaine92}
{Frank}, J., {King}, A., \& {Raine}, D. 1992, {Accretion power in
  astrophysics.}

\bibitem[{{Frank} {et~al.}(1987){Frank}, {King}, \& {Lasota}}]{frank87}
{Frank}, J., {King}, A.~R., \& {Lasota}, J.-P. 1987, A\&A, 178, 137

\bibitem[{{Goranskij} {et~al.}(2007){Goranskij}, {Katysheva}, {Kusakin},
  {Metlova}, {Pogrosheva}, {Shugarov}, {Barsukova}, {Fabrika}, {Borisov},
  {Burenkov}, {Pramsky}, {Karitskaya}, \& {Retter}}]{goranskij07}
{Goranskij}, V.~P., {Katysheva}, N.~A., {Kusakin}, A.~V., {et~al.} 2007,
  Astrophysical Bulletin, 62, 125

\bibitem[{{Greiner}(1996)}]{greiner96}
{Greiner}, J., ed. 1996, Lecture Notes in Physics, Berlin Springer Verlag, Vol.
  472, {Supersoft X-Ray Sources}

\bibitem[{{Greiner} {et~al.}(1991){Greiner}, {Hasinger}, \&
  {Kahabka}}]{sssclass1}
{Greiner}, J., {Hasinger}, G., \& {Kahabka}, P. 1991, A\&A, 246, L17

\bibitem[{{Greiner} {et~al.}(2004){Greiner}, {Iyudin}, {Jimenez-Garate},
  {Burwitz}, {Schwarz}, {DiStefano}, \& {Schulz}}]{greiner_cal87}
{Greiner}, J., {Iyudin}, A., {Jimenez-Garate}, M., {et~al.} 2004, in Revista
  Mexicana de Astronomia y Astrofisica Conference Series, Vol.~20, Revista
  Mexicana de Astronomia y Astrofisica Conference Series, ed. G.~{Tovmassian}
  \& E.~{Sion}, 18

\bibitem[{{Guainazzi} {et~al.}(2009){Guainazzi}, {Risaliti}, {Nucita}, {Wang},
  {Bianchi}, {Soria}, \& {Zezas}}]{ngc1365}
{Guainazzi}, M., {Risaliti}, G., {Nucita}, A., {et~al.} 2009, A\&A, 505, 589

\bibitem[{{Hachisu} {et~al.}(2004){Hachisu}, {Kato}, \& {Kato}}]{hachisu_v1494}
{Hachisu}, I., {Kato}, M., \& {Kato}, T. 2004, ApJL, 606, L139

\bibitem[{{Hachisu} {et~al.}(2000){Hachisu}, {Kato}, {Kato}, {Matsumoto}, \&
  {Nomoto}}]{hachisu00}
{Hachisu}, I., {Kato}, M., {Kato}, T., {Matsumoto}, K., \& {Nomoto}, K. 2000,
  ApJL, 534, L189

\bibitem[{{Hamilton} {et~al.}(2007){Hamilton}, {Urban}, {Sion}, {Riedel},
  {Voyer}, {Marcy}, \& {Lakatos}}]{hamilton07}
{Hamilton}, R.~T., {Urban}, J.~A., {Sion}, E.~M., {et~al.} 2007, ApJ, 667, 1139

\bibitem[{{Hartmann} \& {Heise}(1996)}]{hartheis96}
{Hartmann}, H.~W. \& {Heise}, J. 1996, in Lecture Notes in Physics, Berlin
  Springer Verlag, Vol. 472, Supersoft X-Ray Sources, ed. J.~{Greiner}, 25

\bibitem[{{Hartmann} \& {Heise}(1997)}]{hartheis97}
{Hartmann}, H.~W. \& {Heise}, J. 1997, A\&A, 322, 591

\bibitem[{{Hauschildt} {et~al.}(1992){Hauschildt}, {Wehrse}, {Starrfield}, \&
  {Shaviv}}]{hauschildt92}
{Hauschildt}, P.~H., {Wehrse}, R., {Starrfield}, S., \& {Shaviv}, G. 1992, ApJ,
  393, 307

\bibitem[{{Helton} {et~al.}(2008){Helton}, {Woodward}, {Vanlandingham}, \&
  {Schwarz}}]{helton08}
{Helton}, L.~A., {Woodward}, C.~E., {Vanlandingham}, K., \& {Schwarz}, G.~J.
  2008, CBET, 1379, 1

\bibitem[{{Hounsell} {et~al.}(2010){Hounsell}, {Bode}, {Hick}, {Buffington},
  {Jackson}, {Clover}, {Shafter}, {Darnley}, {Mawson}, {Steele}, {Evans},
  {Eyres}, \& {O'Brien}}]{hounsell10}
{Hounsell}, R., {Bode}, M.~F., {Hick}, P.~P., {et~al.} 2010, ApJ, 724, 480

\bibitem[{{Hutchings} {et~al.}(2002){Hutchings}, {Winter}, {Cowley},
  {Schmidtke}, \& {Crampton}}]{2002AJ....124.2833H}
{Hutchings}, J.~B., {Winter}, K., {Cowley}, A.~P., {Schmidtke}, P.~C., \&
  {Crampton}, D. 2002, AJ, 124, 2833

\bibitem[{{Iijima} \& {Esenoglu}(2003)}]{IE03}
{Iijima}, T. \& {Esenoglu}, H.~H. 2003, A\&A, 404, 997

\bibitem[{{Kahabka}(1996)}]{kahab96}
{Kahabka}, P. 1996, A\&A, 306, 795

\bibitem[{{Kahabka}(1997)}]{cal83_firstoff}
{Kahabka}, P. 1997, in Astronomical Society of the Pacific Conference Series,
  Vol. 121, IAU Colloq. 163: Accretion Phenomena and Related Outflows, ed.
  D.~T. {Wickramasinghe}, G.~V. {Bicknell}, \& L.~{Ferrario}, 730

\bibitem[{{Kahabka} {et~al.}(1994){Kahabka}, {Pietsch}, \&
  {Hasinger}}]{sssclass}
{Kahabka}, P., {Pietsch}, W., \& {Hasinger}, G. 1994, A\&A, 288, 538

\bibitem[{{Kahabka} \& {van den Heuvel}(1997)}]{kahab}
{Kahabka}, P. \& {van den Heuvel}, E.~P.~J. 1997, ARA\&A, 35, 69

\bibitem[{{Kallman} {et~al.}(2003){Kallman}, {Angelini}, {Boroson}, \&
  {Cottam}}]{kallman03}
{Kallman}, T.~R., {Angelini}, L., {Boroson}, B., \& {Cottam}, J. 2003, ApJ,
  583, 861

\bibitem[{{Kaspi} {et~al.}(2002){Kaspi}, {Brandt}, {George}, {Netzer},
  {Crenshaw}, {Gabel}, {Hamann}, {Kaiser}, {Koratkar}, {Kraemer}, {Kriss},
  {Mathur}, {Mushotzky}, {Nandra}, {Peterson}, {Shields}, {Turner}, \&
  {Zheng}}]{ngc3783}
{Kaspi}, S., {Brandt}, W.~N., {George}, I.~M., {et~al.} 2002, ApJ, 574, 643

\bibitem[{{Kinkhabwala} {et~al.}(2002){Kinkhabwala}, {Sako}, {Behar}, {Kahn},
  {Paerels}, {Brinkman}, {Kaastra}, {Gu}, \& {Liedahl}}]{ngc1068}
{Kinkhabwala}, A., {Sako}, M., {Behar}, E., {et~al.} 2002, ApJ, 575, 732

\bibitem[{{Krautter} {et~al.}(1996){Krautter}, {\"Ogelman}, {Starrfield},
  {Wichmann}, \& {Pfeffermann}}]{krautt96}
{Krautter}, J., {\"Ogelman}, H., {Starrfield}, S., {Wichmann}, R., \&
  {Pfeffermann}, E. 1996, ApJ, 456, 788

\bibitem[{{Kuulkers} {et~al.}(2006){Kuulkers}, {Norton}, {Schwope}, \&
  {Warner}}]{kuulkers_review}
{Kuulkers}, E., {Norton}, A., {Schwope}, A., \& {Warner}, B. 2006, {X-rays from
  cataclysmic variables}, ed. W.~H.~G. Lewin \& M.~van~der Klis (Cambridge
  Astrophysics Series, No.~39, Cambridge University Press), 421

\bibitem[{{Lanz} {et~al.}(2005){Lanz}, {Telis}, {Audard}, {Paerels},
  {Rasmussen}, \& {Hubeny}}]{lanz04}
{Lanz}, T., {Telis}, G.~A., {Audard}, M., {et~al.} 2005, ApJ, 619, 517

\bibitem[{{Long} {et~al.}(1981){Long}, {Helfand}, \&
  {Grabelsky}}]{cal_discovery}
{Long}, K.~S., {Helfand}, D.~J., \& {Grabelsky}, D.~A. 1981, ApJ, 248, 925

\bibitem[{{Lyke} \& {Campbell}(2009)}]{v723cas_incl}
{Lyke}, J.~E. \& {Campbell}, R.~D. 2009, AJ, 138, 1090

\bibitem[{{Macri} {et~al.}(2006){Macri}, {Stanek}, {Bersier}, {Greenhill}, \&
  {Reid}}]{MSB06}
{Macri}, L.~M., {Stanek}, K.~Z., {Bersier}, D., {Greenhill}, L.~J., \& {Reid},
  M.~J. 2006, ApJ, 652, 1133

\bibitem[{{Mauche}(2004)}]{mauche04}
{Mauche}, C.~W. 2004, in Revista Mexicana de Astronomia y Astrofisica, vol. 27,
  Vol.~20, Revista Mexicana de Astronomia y Astrofisica Conference Series, ed.
  G.~{Tovmassian} \& E.~{Sion}, 174

\bibitem[{{Mauche} \& {Raymond}(2000)}]{MaucheRaymond2000}
{Mauche}, C.~W. \& {Raymond}, J.~C. 2000, \apj, 541, 924

\bibitem[{{McGowan} {et~al.}(2005){McGowan}, {Charles}, {Blustin}, {Livio},
  {O'Donoghue}, \& {Heathcote}}]{mcgowan05}
{McGowan}, K.~E., {Charles}, P.~A., {Blustin}, A.~J., {et~al.} 2005, MNRAS,
  364, 462

\bibitem[{{Morgan} {et~al.}(2003){Morgan}, {Ringwald}, \& {Prigge}}]{morgan03}
{Morgan}, G.~E., {Ringwald}, F.~A., \& {Prigge}, J.~W. 2003, MNRAS, 344, 521

\bibitem[{{Mukai} \& {Ishida}(2001)}]{mukai01}
{Mukai}, K. \& {Ishida}, M. 2001, ApJ, 551, 1024

\bibitem[{{Munari} {et~al.}(2010){Munari}, {Dallaporta}, \&
  {Castellani}}]{munari10}
{Munari}, U., {Dallaporta}, S., \& {Castellani}, F. 2010, Information Bulletin
  on Variable Stars, 5930, 1

\bibitem[{{Munari} {et~al.}(2013){Munari}, {Dallaporta}, {Castellani},
  {Valisa}, {Frigo}, {Chomiuk}, \& {Ribeiro}}]{2013arXiv1306.1501M}
{Munari}, U., {Dallaporta}, S., {Castellani}, F., {et~al.} 2013, ArXiv e-prints

\bibitem[{{Nelson} {et~al.}(2008){Nelson}, {Orio}, {Cassinelli}, {Still},
  {Leibowitz}, \& {Mucciarelli}}]{nelson07}
{Nelson}, T., {Orio}, M., {Cassinelli}, J.~P., {et~al.} 2008, ApJ, 673, 1067

\bibitem[{{Ness}(2010)}]{ness09}
{Ness}, J. 2010, Astronomische Nachrichten, 331, 179

\bibitem[{{Ness} {et~al.}(2009{\natexlab{a}}){Ness}, {Drake}, {Beardmore},
  {Boyd}, {Bode}, {Brady}, {Evans}, {Gaensicke}, {Kitamoto}, {Knigge},
  {Miller}, {Osborne}, {Page}, {Rodriguez-Gil}, {Schwarz}, {Staels}, {Steeghs},
  {Takei}, {Tsujimoto}, {Wesson}, \& {Zijlstra}}]{v458}
{Ness}, J., {Drake}, J.~J., {Beardmore}, A.~P., {et~al.} 2009{\natexlab{a}},
  AJ, 137, 4160

\bibitem[{{Ness} {et~al.}(2009{\natexlab{b}}){Ness}, {Drake}, {Starrfield},
  {Bode}, {O'Brien}, {Evans}, {Eyres}, {Helton}, {Osborne}, {Page},
  {Schneider}, \& {Woodward}}]{rsophshock}
{Ness}, J., {Drake}, J.~J., {Starrfield}, S., {et~al.} 2009{\natexlab{b}}, AJ,
  137, 3414

\bibitem[{{Ness} {et~al.}(2011){Ness}, {Osborne}, {Dobrotka}, {Page}, {Drake},
  {Pinto}, {Detmers}, {Schwarz}, {Bode}, {Beardmore}, {Starrfield}, {Hernanz},
  {Sala}, {Krautter}, \& {Woodward}}]{ness_v2491}
{Ness}, J., {Osborne}, J.~P., {Dobrotka}, A., {et~al.} 2011, ApJ, 733, 70

\bibitem[{{Ness} {et~al.}(2012{\natexlab{a}}){Ness}, {Schaefer}, {Dobrotka},
  {Sadowski}, {Drake}, {Barnard}, {Talavera}, {Gonzalez-Riestra}, {Page},
  {Hernanz}, {Sala}, \& {Starrfield}}]{ness_usco}
{Ness}, J., {Schaefer}, B.~E., {Dobrotka}, A., {et~al.} 2012{\natexlab{a}},
  ApJ, 745, 43

\bibitem[{{Ness} {et~al.}(2007{\natexlab{a}}){Ness}, {Schwarz}, {Retter},
  {Starrfield}, {Schmitt}, {Gehrels}, {Burrows}, \& {Osborne}}]{swnovae}
{Ness}, J., {Schwarz}, G.~J., {Retter}, A., {et~al.} 2007{\natexlab{a}}, ApJ,
  663, 505

\bibitem[{{Ness} {et~al.}(2007{\natexlab{b}}){Ness}, {Starrfield}, {Beardmore},
  {Bode}, {Drake}, {Evans}, {Gehrz}, {Goad}, {Gonzalez-Riestra}, {Hauschildt},
  {Krautter}, {O'Brien}, {Osborne}, {Page}, {Sch\"onrich}, \&
  {Woodward}}]{ness_rsoph}
{Ness}, J., {Starrfield}, S., {Beardmore}, A., {et~al.} 2007{\natexlab{b}},
  ApJ, 665, 1334

\bibitem[{{Ness} {et~al.}(2003){Ness}, {Starrfield}, {Burwitz}, {Wichmann},
  {Hauschildt}, {Drake}, {Wagner}, {Bond}, {Krautter}, {Orio}, {Hernanz},
  {Gehrz}, {Woodward}, {Butt}, {Mukai}, {Balman}, \& {Truran}}]{v4743}
{Ness}, J., {Starrfield}, S., {Burwitz}, V., {et~al.} 2003, ApJL, 594, L127

\bibitem[{{Ness} {et~al.}(2005){Ness}, {Starrfield}, {Jordan}, {Krautter}, \&
  {Schmitt}}]{ness_vel}
{Ness}, J., {Starrfield}, S., {Jordan}, C., {Krautter}, J., \& {Schmitt},
  J.~H.~M.~M. 2005, MNRAS, 364, 1015

\bibitem[{{Ness}(2012)}]{basi}
{Ness}, J.~U. 2012, Bulletin of the Astronomical Society of India, 40, 353

\bibitem[{{Ness} {et~al.}(2012{\natexlab{b}}){Ness}, {Shore}, {Drake},
  {Osborne}, {Page}, {Beardmore}, {Schwarz}, \& {Starrfield}}]{ATel4569}
{Ness}, J.-U., {Shore}, S.~N., {Drake}, J.~J., {et~al.} 2012{\natexlab{b}}, The
  Astronomer's Telegram, 4569, 1

\bibitem[{{Orio} {et~al.}(2013){Orio}, {Behar}, {Gallagher}, {Bianchini},
  {Chiosi}, {Luna}, {Nelson}, {Rauch}, {Schaefer}, \& {Tofflemire}}]{orio_usco}
{Orio}, M., {Behar}, E., {Gallagher}, J., {et~al.} 2013, MNRAS, 429, 1342

\bibitem[{{Orio} {et~al.}(2001){Orio}, {Parmar}, {Benjamin}, {Amati},
  {Frontera}, {Greiner}, {{\" O}gelman}, {Mineo}, {Starrfield}, \&
  {Trussoni}}]{Orio2001}
{Orio}, M., {Parmar}, A., {Benjamin}, R., {et~al.} 2001, MNRAS, 326, L13

\bibitem[{{Orio} {et~al.}(2002){Orio}, {Parmar}, {Greiner}, {{\" O}gelman},
  {Starrfield}, \& {Trussoni}}]{orio02}
{Orio}, M., {Parmar}, A.~N., {Greiner}, J., {et~al.} 2002, MNRAS, 333, L11

\bibitem[{{Osborne} {et~al.}(2011){Osborne}, {Page}, {Beardmore}, {Bode},
  {Goad}, {O'Brien}, {Starrfield}, {Rauch}, {Ness}, {Krautter}, {Schwarz},
  {Burrows}, {Gehrels}, {Drake}, {Evans}, \& {Eyres}}]{osborne11}
{Osborne}, J.~P., {Page}, K.~L., {Beardmore}, A.~P., {et~al.} 2011, ApJ, 727,
  124

\bibitem[{{Page} {et~al.}(2010){Page}, {Osborne}, {Evans}, {Wynn}, {Beardmore},
  {Starling}, {Bode}, {Ibarra}, {Kuulkers}, {Ness}, \& {Schwarz}}]{page09}
{Page}, K.~L., {Osborne}, J.~P., {Evans}, P.~A., {et~al.} 2010, MNRAS, 401, 121

\bibitem[{{Page} {et~al.}(2013){Page}, {Osborne}, {Wagner}, {Beardmore},
  {Shore}, {Starrfield}, \& {Woodward}}]{page_mon2012}
{Page}, K.~L., {Osborne}, J.~P., {Wagner}, R.~M., {et~al.} 2013, ApJL, 768, L26

\bibitem[{{Parmar} {et~al.}(1997){Parmar}, {Kahabka}, {Hartmann}, {Heise},
  {Martin}, {Bavdaz}, \& {Mineo}}]{parmarcal87}
{Parmar}, A.~N., {Kahabka}, P., {Hartmann}, H.~W., {et~al.} 1997, A\&A, 323,
  L33

\bibitem[{{Parmar} {et~al.}(1998){Parmar}, {Kahabka}, {Hartmann}, {Heise}, \&
  {Taylor}}]{parmarcal83}
{Parmar}, A.~N., {Kahabka}, P., {Hartmann}, H.~W., {Heise}, J., \& {Taylor},
  B.~G. 1998, A\&A, 332, 199

\bibitem[{{Ragan} {et~al.}(2009){Ragan}, {Brozek}, {Suchomska}, {Skalbania},
  {Konorski}, {Galan}, {Swierczynski}, {Tomov}, {Mikolajewski}, \&
  {Wychudzki}}]{RBS09}
{Ragan}, E., {Brozek}, T., {Suchomska}, K., {et~al.} 2009, The Astronomer's
  Telegram, 2327, 1

\bibitem[{{Rajoelimanana} {et~al.}(2013){Rajoelimanana}, {Charles}, {Meintjes},
  {Odendaal}, \& {Udalski}}]{cal83_optcorr}
{Rajoelimanana}, A.~F., {Charles}, P.~A., {Meintjes}, P.~J., {Odendaal}, A., \&
  {Udalski}, A. 2013, MNRAS, 432, 2886

\bibitem[{{Rauch} {et~al.}(2010){Rauch}, {Orio}, {Gonzales-Riestra}, {Nelson},
  {Still}, {Werner}, \& {Wilms}}]{rauch10}
{Rauch}, T., {Orio}, M., {Gonzales-Riestra}, R., {et~al.} 2010, ApJ, 717, 363

\bibitem[{{Ribeiro} {et~al.}(2013{\natexlab{a}}){Ribeiro}, {Bode}, {Darnley},
  {Barnsley}, {Munari}, \& {Harman}}]{RBD13}
{Ribeiro}, V.~A.~R.~M., {Bode}, M.~F., {Darnley}, M.~J., {et~al.}
  2013{\natexlab{a}}, MNRAS, 433, 1991

\bibitem[{{Ribeiro} {et~al.}(2009){Ribeiro}, {Bode}, {Darnley}, {Harman},
  {Newsam}, {O'Brien}, {Bohigas}, {Echevarr{\'{\i}}a}, {Bond}, {Chavushyan},
  {Costero}, {Coziol}, {Evans}, {Eyres}, {Le{\'o}n-Tavares}, {Richer},
  {Tovmassian}, {Starrfield}, \& {Zharikov}}]{ribeiro09}
{Ribeiro}, V.~A.~R.~M., {Bode}, M.~F., {Darnley}, M.~J., {et~al.} 2009, \apj,
  703, 1955

\bibitem[{{Ribeiro} {et~al.}(2011){Ribeiro}, {Darnley}, {Bode}, {Munari},
  {Harman}, {Steele}, \& {Meaburn}}]{ribeiro10}
{Ribeiro}, V.~A.~R.~M., {Darnley}, M.~J., {Bode}, M.~F., {et~al.} 2011, MNRAS,
  412, 1701

\bibitem[{{Ribeiro} {et~al.}(2013{\natexlab{b}}){Ribeiro}, {Munari}, \&
  {Valisa}}]{ribeiro13}
{Ribeiro}, V.~A.~R.~M., {Munari}, U., \& {Valisa}, P. 2013{\natexlab{b}}, ApJ,
  768, 49

\bibitem[{{Rohrbach} {et~al.}(2009){Rohrbach}, {Ness}, \&
  {Starrfield}}]{rohrbach09}
{Rohrbach}, J.~G., {Ness}, J., \& {Starrfield}, S. 2009, AJ, 137, 4627

\bibitem[{{Sala} {et~al.}(2008){Sala}, {Hernanz}, {Ferri}, \&
  {Greiner}}]{sala08}
{Sala}, G., {Hernanz}, M., {Ferri}, C., \& {Greiner}, J. 2008, ApJL, 675, L93

\bibitem[{{Sala} {et~al.}(2010){Sala}, {Hernanz}, {Ferri}, \&
  {Greiner}}]{sala10}
{Sala}, G., {Hernanz}, M., {Ferri}, C., \& {Greiner}, J. 2010, Astronomische
  Nachrichten, 331, 201

\bibitem[{{Schaefer}(1990)}]{schaefer_eclipse}
{Schaefer}, B.~E. 1990, ApJL, 355, L39

\bibitem[{{Schaefer}(2010)}]{schaefer10}
{Schaefer}, B.~E. 2010, ApJS, 187, 275

\bibitem[{{Schaeidt} {et~al.}(1993){Schaeidt}, {Hasinger}, \&
  {Truemper}}]{rxj_onoff}
{Schaeidt}, S., {Hasinger}, G., \& {Truemper}, J. 1993, A\&A, 270, L9

\bibitem[{{Schandl} {et~al.}(1997){Schandl}, {Meyer-Hofmeister}, \&
  {Meyer}}]{schandl97}
{Schandl}, S., {Meyer-Hofmeister}, E., \& {Meyer}, F. 1997, A\&A, 318, 73

\bibitem[{{Schlegel} {et~al.}(2010){Schlegel}, {Schaefer}, {Pagnotta}, {Page},
  {Osborne}, {Drake}, {Orio}, {Takei}, {Kuulkers}, \& {Ness}}]{ATel2419}
{Schlegel}, E.~M., {Schaefer}, B., {Pagnotta}, A., {et~al.} 2010, The
  Astronomer's Telegram, 2419, 1

\bibitem[{{Schmidtke} \& {Cowley}(2006)}]{schmidtke06}
{Schmidtke}, P.~C. \& {Cowley}, A.~P. 2006, AJ, 131, 600

\bibitem[{{Sch{\"o}nrich} \& {Ness}(2008)}]{schoenrich07}
{Sch{\"o}nrich}, R.~A. \& {Ness}, J.-U. 2008, in Astronomical Society of the
  Pacific Conference Series, Vol. 401, RS Ophiuchi (2006) and the Recurrent
  Nova Phenomenon, ed. A.~{Evans}, M.~F. {Bode}, T.~J. {O'Brien}, \& M.~J.
  {Darnley}, 291

\bibitem[{{Schwarz} {et~al.}(2011){Schwarz}, {Ness}, {Osborne}, {Page},
  {Evans}, {Beardmore}, {Walter}, {Helton}, {Woodward}, {Bode}, {Starrfield},
  \& {Drake}}]{schwarz2011}
{Schwarz}, G.~J., {Ness}, J., {Osborne}, J.~P., {et~al.} 2011, ApJS, 197, 31

\bibitem[{{Schwarz} {et~al.}(2008){Schwarz}, {Ness}, {Osborne}, {Page},
  {Wagner}, {Starrfield}, {Prieto}, {Pejcha}, \& {Denney}}]{SNO08}
{Schwarz}, G.~J., {Ness}, J.-U., {Osborne}, J.~P., {et~al.} 2008, The
  Astronomer's Telegram, 1847, 1

\bibitem[{{Shore} {et~al.}(2011){Shore}, {Augusteijn}, {Ederoclite}, \&
  {Uthas}}]{2011A&A...533L...8S}
{Shore}, S.~N., {Augusteijn}, T., {Ederoclite}, A., \& {Uthas}, H. 2011, A\&A,
  533, L8

\bibitem[{{Shore} {et~al.}(2003){Shore}, {Schwarz}, {Bond}, {Downes},
  {Starrfield}, {Evans}, {Gehrz}, {Hauschildt}, {Krautter}, \&
  {Woodward}}]{shore03}
{Shore}, S.~N., {Schwarz}, G., {Bond}, H.~E., {et~al.} 2003, AJ, 125, 1507

\bibitem[{{Sokoloski} {et~al.}(2013){Sokoloski}, {Crotts}, {Lawrence}, \&
  {Uthas}}]{SCL13}
{Sokoloski}, J.~L., {Crotts}, A.~P.~S., {Lawrence}, S., \& {Uthas}, H. 2013,
  ApJl, 770, L33

\bibitem[{{Southwell} {et~al.}(1996){Southwell}, {Livio}, {Charles},
  {O'Donoghue}, \& {Sutherland}}]{southwell96}
{Southwell}, K.~A., {Livio}, M., {Charles}, P.~A., {O'Donoghue}, D., \&
  {Sutherland}, W.~J. 1996, ApJ, 470, 1065

\bibitem[{{Starrfield} {et~al.}(2008){Starrfield}, {Iliadis}, \& {Hix}}]{st08}
{Starrfield}, S., {Iliadis}, C., \& {Hix}, W.~R. 2008, in Classical Novae, ed.
  M.~Bode \& A.~Evans (Cambridge University Press), 77

\bibitem[{{Suleimanov} {et~al.}(1999){Suleimanov}, {Meyer}, \&
  {Meyer-Hofmeister}}]{suleimanov99}
{Suleimanov}, V., {Meyer}, F., \& {Meyer-Hofmeister}, E. 1999, A\&A, 350, 63

\bibitem[{{Suleimanov} {et~al.}(2003){Suleimanov}, {Meyer}, \&
  {Meyer-Hofmeister}}]{suleimanov03}
{Suleimanov}, V., {Meyer}, F., \& {Meyer-Hofmeister}, E. 2003, A\&A, 401, 1009

\bibitem[{{Takei} {et~al.}(2012){Takei}, {Drake}, {Ness}, {Starrfield},
  {Schwarz}, {Page}, {Osborne}, {Rossum}, \& {Walter}}]{LMC2012_chan}
{Takei}, D., {Drake}, J.~J., {Ness}, J.-U., {et~al.} 2012, The Astronomer's
  Telegram, 4116, 1

\bibitem[{{Thoroughgood} {et~al.}(2001){Thoroughgood}, {Dhillon}, {Littlefair},
  {Marsh}, \& {Smith}}]{thoroughgood01}
{Thoroughgood}, T.~D., {Dhillon}, V.~S., {Littlefair}, S.~P., {Marsh}, T.~R.,
  \& {Smith}, D.~A. 2001, MNRAS, 327, 1323

\bibitem[{{Tofflemire} {et~al.}(2011){Tofflemire}, {Orio}, {Kuulkers},
  {Osborne}, {Page}, {Beardmore}, {Drake}, {Ness}, {Shore}, \&
  {Starrfield}}]{tpyx_chan}
{Tofflemire}, B., {Orio}, M., {Kuulkers}, E., {et~al.} 2011, The Astronomer's
  Telegram, 3762, 1

\bibitem[{{Tomov} {et~al.}(1998){Tomov}, {Munari}, {Kolev}, {Tomasella}, \&
  {Rejkuba}}]{tomov98}
{Tomov}, T., {Munari}, U., {Kolev}, D., {Tomasella}, L., \& {Rejkuba}, M. 1998,
  A\&A, 333, L67

\bibitem[{{Truemper}(1992)}]{truemper92}
{Truemper}, J. 1992, QJRAS, 33, 165

\bibitem[{{Uthas} {et~al.}(2010){Uthas}, {Knigge}, \& {Steeghs}}]{tpyx_pole2}
{Uthas}, H., {Knigge}, C., \& {Steeghs}, D. 2010, MNRAS, 409, 237

\bibitem[{{van den Heuvel} {et~al.}(1992){van den Heuvel}, {Bhattacharya},
  {Nomoto}, \& {Rappaport}}]{heuvel}
{van den Heuvel}, E.~P.~J., {Bhattacharya}, D., {Nomoto}, K., \& {Rappaport},
  S.~A. 1992, A\&A, 262, 97

\bibitem[{{van Rossum}(2012)}]{vanRossum2012}
{van Rossum}, D.~R. 2012, ApJ, 756, 43

\bibitem[{{Vanlandingham} {et~al.}(2007){Vanlandingham}, {Schwarz},
  {Starrfield}, {Woodward}, {Wagner}, {Ness}, \& {Helton}}]{VSS07}
{Vanlandingham}, K.~M., {Schwarz}, G., {Starrfield}, S., {et~al.} 2007, in
  Bulletin of the American Astronomical Society, Vol.~39, American Astronomical
  Society Meeting Abstracts \#210, 99

\bibitem[{{Walter} \& {Battisti}(2011)}]{walter11}
{Walter}, F.~M. \& {Battisti}, A. 2011, in Bulletin of the American
  Astronomical Society, Vol.~43, American Astronomical Society Meeting
  Abstracts \#217, \#338.11

\bibitem[{{Worters} {et~al.}(2010){Worters}, {Eyres}, {Rushton}, \&
  {Schaefer}}]{IAUC9114}
{Worters}, H.~L., {Eyres}, S.~P.~S., {Rushton}, M.~T., \& {Schaefer}, B. 2010,
  IAUC, 9114, 1

\end{thebibliography}

\end{document}